\documentclass[journal]{IEEEtran}
\usepackage{cite}

\usepackage{booktabs}
\usepackage{marginnote}
\usepackage{threeparttable}
\usepackage{multirow}
\usepackage[ruled,linesnumbered]{algorithm2e}

\ifCLASSINFOpdf
  \usepackage[pdftex]{graphicx}
  \graphicspath{{./figs/}{./figs_realdata/}}
  \DeclareGraphicsExtensions{.pdf,.jpeg,.png}
\else
  \usepackage[dvips]{graphicx}
  \graphicspath{{../eps/}}
  \DeclareGraphicsExtensions{.eps}
\fi
\usepackage{easyReview}
\usepackage{todonotes}
\usepackage{amssymb}

\makeatletter

\newcommand{\Rmnum}[1]{\expandafter\@slowromancap\romannumeral #1@}
\makeatother

\usepackage{amsmath}

\usepackage{xcolor}

\ifCLASSOPTIONcompsoc
  \usepackage[caption=false,font=normalsize,labelfont=sf,textfont=sf]{subfig}
\else
  \usepackage[caption=false,font=footnotesize]{subfig}
\fi

\usepackage{stfloats}
\begin{document}
%
\title{Prior-Guided Deep Interference Mitigation \\ for FMCW Radars}
%
%
%

\author{Jianping~Wang,~Runlong~Li,~Yuan~He,~and~Yang~Yang
\thanks{J. Wang is with the Faculty
of Electrical Engineering, Mathematics and Computer Science, Delft University of Technology, Delft, 2628CD, the Netherlands. e-mail: J.Wang-4@tudelft.nl.}
\thanks{R. Li and Y. He are with Key Laboratory of Trustworthy Distributed Computing and Service, Beijing University of Posts and Telecommunications, Beijing, 100876, China. email: lirunlong@bupt.edu.cn, yuanhe@bupt.edu.cn.} 
\thanks{Y. Yang is with School of Electrical and Information Engineering, Tianjin University, Tianjin, 300072 China. email: yang\_yang@tju.edu.cn.}

}

\maketitle

\begin{abstract}
A prior-guided deep learning (DL) based interference mitigation approach is proposed for frequency modulated continuous wave (FMCW) radars. In this paper, the interference mitigation problem is tackled as a regression problem. Considering the complex-valued nature of radar signals, the complex-valued convolutional neural network is utilized as an architecture for implementation, which is different from the conventional real-valued counterparts. Meanwhile, as the useful beat signals of FMCW radars and interferences exhibit different distributions in the time-frequency domain, this prior feature is exploited as a regularization term to avoid overfitting of the learned representation. The effectiveness and accuracy of our proposed complex-valued fully convolutional network (CV-FCN) based interference mitigation approach are verified and analyzed through both simulated and measured radar signals. Compared to the real-valued counterparts, the CV-FCN shows a better interference mitigation performance with a potential of half memory reduction in low Signal to Interference plus Noise Ratio (SINR) scenarios. Moreover, the CV-FCN trained using only simulated data can be directly utilized for interference mitigation in various measured radar signals and shows a superior generalization capability. Furthermore, by incorporating the prior feature, the CV-FCN trained on only 1/8 of the full data achieves comparable performance as that on the full dataset in low SINR scenarios, and the training procedure converges faster.
\end{abstract}


%
\IEEEpeerreviewmaketitle

\section{Introduction}
%
%
%
%
\IEEEPARstart{F}{requency} modulated continuous wave (FMCW) radars are widely used for automotive radar, vital sign detection, smart building surveillance, weather monitoring, etc. 
With the rapid expansion of the applications, the mutual interference among FMCW radars as well as surrounding wireless devices becomes an increasingly severe problem, which would mask weak targets, degrade target detection and even cause ghost targets.

A number of methods have been proposed for FMCW radar interference mitigation (IM), including radar system coordination, radar system design and waveform design, and signal processing. 
Radar system coordination can operate at both the transmitter (Tx) and receiver (Rx) end\cite{Autoregressive}, which may introduce an additional communication module in the radar system.
In contrast, at the Tx end, a new radar system or waveform can be designed to transmit the chirp signal with varying parameters (e.g., center frequency) to avoid the appearance of interferences.
Besides, the interference can be suppressed in the RX end by applying the traditional signal processing approaches or the latest deep learning (DL) based approaches to the received radar signals.
The signal post-processing method does not require to design a new radar system and is easier to fit into existing radar chips.

Specifically, the traditional signal processing approaches can be classified into three categories: zeroing and reconstruction, estimation and subtraction, and digital beamforming. 
In \cite{sparse_sampling}, the amplitudes of interfered radar signals are replaced with zero, and the useful beat signals are reconstructed by an iterative method in the Fourier domain. 
However, the reconstruction algorithm would be inapplicable especially in the long interference duration situations. 
By contrast, the parameters of interfering signals can be estimated, and only the interference components are subtracted from the received signals\cite{estimation}, which requires more computational effort and long processing time. 
Meanwhile, an adaptive noise canceller (ANC) utilizing the symmetry of interferences in the frequency domain is used to remove the interferences in the positive frequency with lower computational complexity\cite{ANC}. 
In \cite{Wavelet_Denoising}, the interferences are reconstructed and subtracted by wavelet denoising (WD), whose performance would degrade as the proportion of interferences increases. 
Furthermore, the filtering can also be used for IM, which has a distortion of object peak values due to its non-linear nature\cite{nonlinear_filter}. 
Finally, in multi-antenna systems, the interference from certain directions can be removed in the space domain by digital beamforming\cite{beamforming}. 
However, the targets' signals scattered from the same directions are also suppressed.

The traditional signal processing approaches can effectively suppress the interferences and improve the targets' detection probability for FMCW radars. However, these approaches generally require high computational complexity, and the performance depends on the manually selected parameters. Besides, most of the traditional approaches have made many assumptions to simplify the calculation and obtain analytical solutions, whose performance would significantly degrade in more complex scenarios. 

Recent development in deep learning has shown its ability in feature extraction, and DL-based approaches are increasingly used for various signal processing problems\cite{Human}, including interference mitigation for radar signals. In \cite{CNN,CNN_real,CNN_new}, a simple Convolutional Neural Network (CNN) with few parameters is used to suppress the interference and noise by operating on the range-doppler (RD) maps for FMCW radars. Then more complex network structures including Fully convolutional network (FCN)\cite{FCN}, Autoencoder\cite{Autoencoder} and U-Net\cite{Unet} are proposed to process the frequency spectra or the RD maps of radar signals. These approaches extract the feature of interferences and subtract it from received signals. Similarly, in \cite{resnet}, the CNN and residual network (ResNet) are built to detect and remove the interference components respectively for synthetic aperture radars. Moreover, the Generative Adversarial Network (GAN) can be used to recover the missing signals after interference detection and zeroing\cite{GAN_IM}. Besides, Recurrent Neural Networks (RNNs) are also implemented in the time domain with low processing time \cite{RNN,Attention}. 

Compared to the traditional signal processing methods, the DL-based approaches require building a training dataset and designing the neural network architecture, which can automatically extract the critical features through training, distinguishing the targets' scattered signals and interferences.
The experimental results have shown its powerful interference mitigation ability. Moreover, the DL-based approaches can apply in more complex situations by learning a causal model from data instead of building a specific signal model. 
On the other hand, due to the limitation of a large-scale dataset including radar signals in various scenarios, it is hard to acquire satisfactory results for DL-based approaches.
Additionally, with the existence of the overfitting problem, the features extracted by the network may be affected by noise. As a result, the performance of existing DL-based approaches is limited by the number of radar signals that are difficult to collect, and the total parameters may exceed the capacity of existing small memory-constrained radar sensors.

Generally, the beat signals of FMCW radars are acquired as complex-valued samples with I/Q receivers. The existing DL-based IM approaches all separate the complex-valued samples as real and imaginary parts and handle them as independent real-valued data with real-valued neural networks. Thus the implicit relationship between the real and imaginary parts of radar signals is not considered, which may cause the loss of the phase information that is necessary for further signal processing steps, for instance, classification and tracking. 
On the other hand, complex-valued convolutional neural network \cite{complex_classification,complex_mri,complex_autoencoder}, which handles complex-valued data with the algebraic rules of complex numbers, can achieve better performance than the real-valued counterparts. Moreover, the complex-valued network has a more powerful representation ability and is robust to noise\cite{complex}. Its potential for faster learning, easier optimization, and better generalization performance has received increasing attention in various domains\cite{hirose2012generalization}.

\begin{figure*}[!b]
	\centering
	\includegraphics[width=\textwidth]{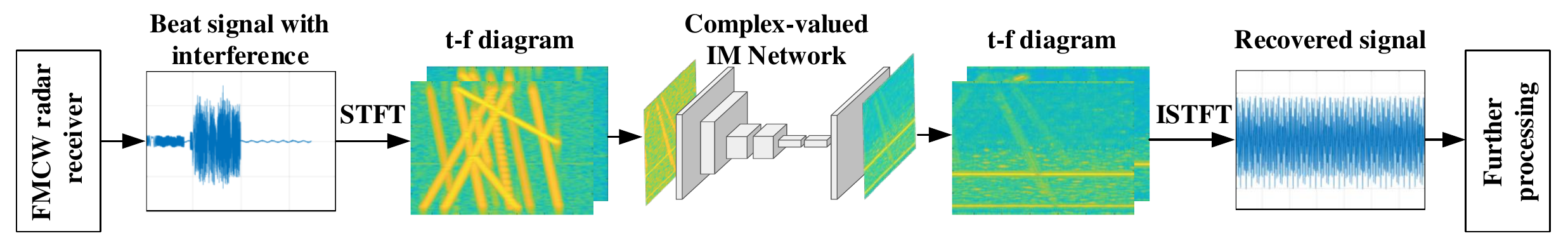}
	\caption{Signal processing flow of our propose approach for interference mitigation (IM).}
	\label{fig:signal_processing_flow}
\end{figure*}

In this paper, considering the complex-valued nature of beat signals of FMCW radars, an interference mitigation approach based on complex-valued convolutional neural networks is proposed (see Fig. \ref{fig:signal_processing_flow}). Specifically, a complex-valued fully convolutional network (CV-FCN) is designed to suppress the interference and noise, which operates on the spectrogram in the time-frequency ($t$-$f$) domain obtained by taking a short-time Fourier transform (STFT) of a beat signal.
Moreover, accounting for that the beat signals are shown as straight lines parallel to the time axis in the spectrogram while interferences exhibit as oblique lines in the $t$-$f$ domain, we exploit this prior feature as a regularization term combined with the mean square error (MSE) loss function for training.

The contributions of this paper are summarized as follows:

Firstly, a CV-FCN-based approach is proposed for interference mitigation. 
By using complex convolutions and the activation function CReLU, a better interference mitigation performance with a potential of half memory reduction compared to the real-valued counterparts in low SINR scenarios is achieved. 

Secondly, a prior-guided loss function is proposed by accounting for both data consistency between labels and the predicted $t$-$f$ spectra and the expected prior frequency-sparse feature of the predicted spectra. A hyper-parameter is used to trade off between the data consistency and the expected prior feature of the predicted spectra. By adjusting the hyper-parameter, the overfitting problem can be avoided, and the networks can be trained with a smaller dataset and faster convergence.

Thirdly, the proposed approach to IM can process radar signals with an arbitrary length in a sweep. 
Its performance is verified through both simulated and measured data, showing its effectiveness and superior generalization capability.

The remainder of this paper is organized as follows. Section II introduces the signal modelling and analysis. Section III elaborates the prior-guided interference mitigation approach based on complex-valued convolutional neural networks. The setups of numerical simulations for data synthesis and experimental measurements are described in Section IV. After that, the experimental results of the proposed method on the simulated and measured radar signals are presented in Sections V and VI. Finally, conclusions are drawn in Section VII.

\section{Signal Modelling and Analysis}
Dechirping receiver is widely used for FMCW radar system to reduce the sampling requirement to the analog to digital converters (ADC). After dechirping, the acquired beat signals contaminated by interferences can be written as \cite{wang2021matrix}:
\begin{equation}
    y(t) = s(t) + f(t) + n(t)
 	\label{equa_1}
\end{equation}
where $n(t)$ represents the thermal noise and measurement errors. $s(t)$ is the useful beat signals and $f(t)$ denotes the interferences, which are explicitly given by 
\begin{align}
s(t) &= \sum_{k=1}^{N} \sigma_k \exp\left[j2\pi \left(-f_c \tau_k - K\tau_k t + \frac{1}{2} K \tau_k ^2 \right) \right]   \\
f(t) &= \mathcal{F}_{lp}\left[p^\ast(t)\sum_{m=1}^{M}f_m(t) \right]
\end{align}
where $0<t<T_{\text{sw}}$ with sweep duration $T_{\text{sw}}$. $f_c$ is the center frequency, $K$ is the chirp rate of the FMCW waveform, and $\tau_k$ is the time delay of the scattered signal from the $k^\text{th}$ target relative to the transmitted one. $p^\ast(t)$ is the reference signal used for dechirping, $f_m(t)$ denotes the $m^\text{th}$ interference and $\mathcal{F}_{lp}$ is the low-pass filtering (LPF) operator. 

\begin{figure}[!t]
	\centering
	\includegraphics[width=0.35\textwidth]{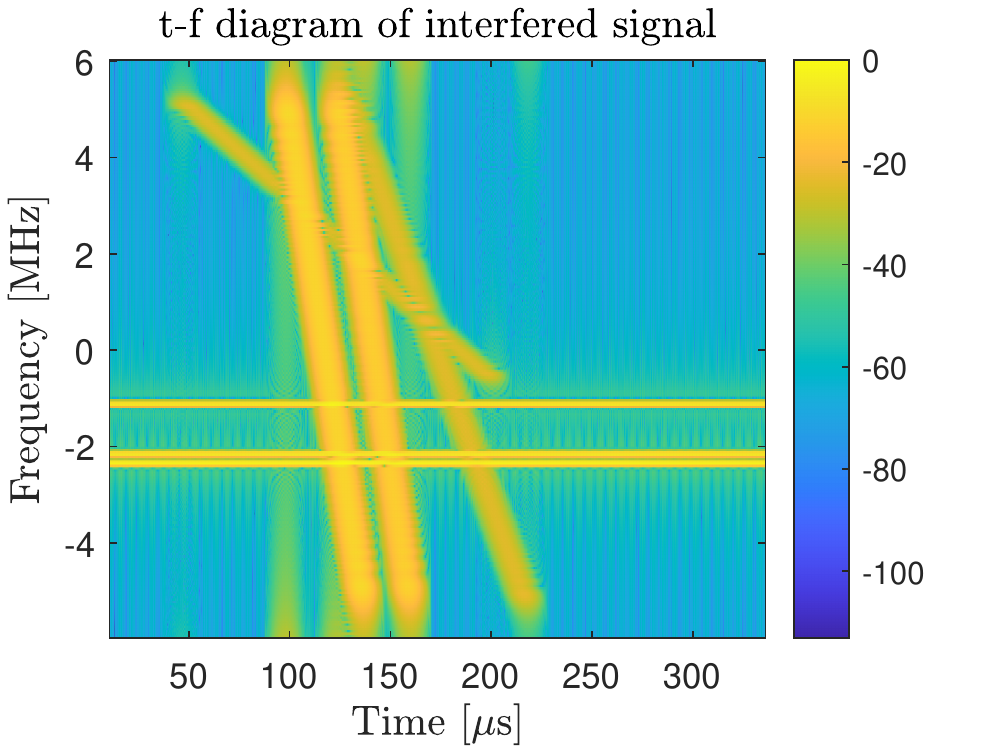}
	\caption{$t$-$f$ diagram of beat signals contaminated by mutual interferences.}
	\label{fig:stft}
\end{figure}

In practice, the interference component $f(t)$ could result from aggressor FMCW radars or other neighboring wireless devices. As analyzed in \cite{wang2021matrix}, the interferences can be, generally classified as four categories: (i) interference signal with the same chirp rate; (ii) interference signal with different chirp rate; (iii) CW interference; and (iv) transient interference. In case (i), when the beat signals resulting from interferences fall in the effective bandwidth of the LPF, it would result in horizontal lines along the time axis same as that of real targets' signals in the $t$-$f$ spectrum; thus, it would cause ghost targets, degrading the probability of detection and the false alarm rate. It is difficult to mitigate this kind of interference in the time or frequency domain, and this problem may be solved by designing a specific radar system or waveform in the space domain. In cases (ii)-(iv), the interferences lead to inclined thick lines or superposition of inclined thick lines in the $t$-$f$ spectrum, which are different from the beat signals of targets. Therefore, without loss of generality, we consider interferences in case (ii) for demonstration in this paper. Note the approach is also applicable to handle interferences in cases (iii) and (iv).

Fig. \ref{fig:stft} shows the $t$-$f$ spectrum of an interference-contaminated beat signal, where the horizontal lines along the time axis are the spectra of targets' signal components while the inclined thick lines are the interferences. The interferences show different distributions determined by their amplitudes, chirp rates, duration times, and time delays relative to the reference signal for dechirping. Moreover, due to the positive time delay caused by wave propagation, the spectra of beat signals always exist in negative frequency (the sweep slope of the victim radar is positive). In contrast, the spectra of interferences spread in both positive and negative frequency in the $t$-$f$ spectrum. 
Considering the different temporal and spectral features of targets' signals and interferences, it is natural to investigate possible approaches to mitigate the interferences in the $t$-$f$ domain by processing, for instance, the STFT spectrum. 

Based on the differences, the interference mitigation problem can be tackled as a two-step interference detection and suppression problem by using the positive-frequency spectrum to detect if interferences exist or not. Besides, the recent development in deep learning techniques substantially improves the detection performance by exploiting multi-layer CNN. As a result, this problem is tackled as a single-stage regression problem based on deep learning in this paper, which means detecting and mitigating the interferences can be completed by only an end-to-end neural network.

\section{Prior-Guided Deep Interference Mitigation}

In this section, some basic modules used in the complex-valued convolutional neural network are first reviewed. 
Then the complex-valued fully convolutional network architecture and the prior-guided loss function proposed for FMCW radar interference mitigation are introduced, followed by the detailed description of the training procedure.

\subsection{Complex-valued modules}

A complex-valued convolutional neural network is generally composed of various complex-valued modules, including complex convolution, complex-valued activation functions and complex batch normalization \cite{complex}.

To take advantage of the existing deep learning platform developed for real-valued NN (e.g., TensorFlow \cite{abadi2016tensorflow}), a complex convolution can be implemented by explicitly performing real-valued convolutions among the real and imaginary parts of the related terms. 
Specifically, the complex convolution between a complex filter $\mathbf{W}=\mathbf{A}+j\mathbf{B}$ and a complex vector $\mathbf{h}=\mathbf{x} +j\mathbf{y}$ can be expressed as:
\begin{equation}
	\mathbf{W}*\mathbf{h} = (\mathbf{A}*\mathbf{x}-\mathbf{B}*\mathbf{y}) + j(\mathbf{A}*\mathbf{y}+\mathbf{B}*\mathbf{x})
	\label{equa_2}
\end{equation}
where $j$ is the imaginary unit. $\mathbf{A}$ and $\mathbf{B}$ are real matrices, and $\mathbf{x}$ and $\mathbf{y}$ are real vectors, respectively. 

Similar to the activation functions for real-valued CNNs, complex-valued activation functions introduce nonlinearity to complex-valued CNNs to increase their representation capabilities \cite{complex,modrelu,zrelu}. 
Complex Rectified Linear Unit (CReLU) is one of the most popular activation functions used in complex-valued CNNs \cite{complex}, which applies traditional real-valued ReLU on both real and imaginary parts of a complex-valued input and is expressed as:
\begin{equation}
	\text{CReLU}(z) = \text{ReLU}( \Re(z)) + j \text{ReLU}(\Im(z))
	\label{equa_4}
\end{equation}
where $\Re(x)$ and $\Im(x)$ extract the real and imaginary parts of a complex number $x$. 
Compared to other complex-valued activation functions (e.g., zReLU \cite{zrelu} and modReLU \cite{modrelu}), CReLU generally achieves the best performance in inverse problems. Therefore, it is utilized in this paper as well. 

\subsection{Network Architecture} 
\begin{figure*}[!t]
	\centering
	\vspace{-3mm}
	\includegraphics[width=0.95\textwidth]{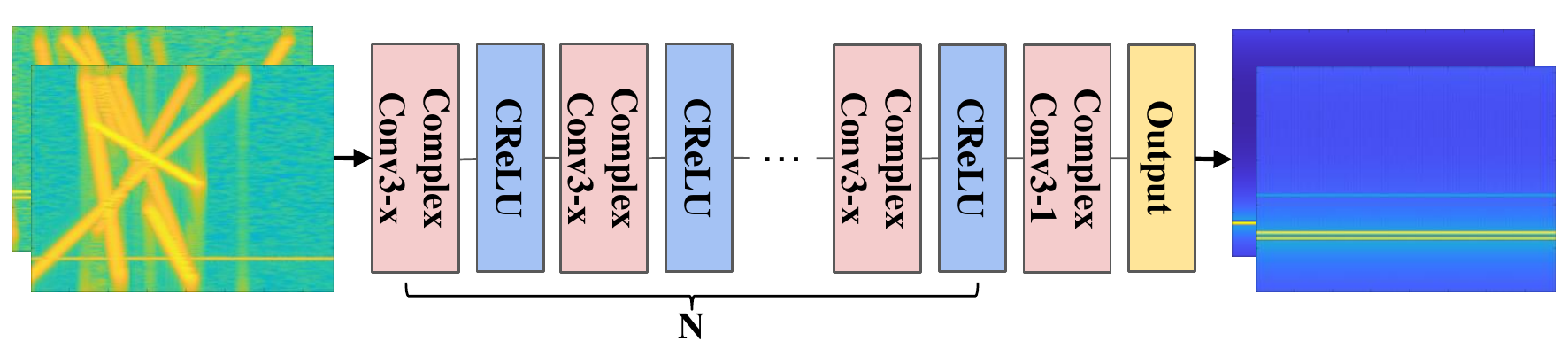}
	\caption{Proposed complex-valued fully convolutional network architecture (CV-FCN). It uses complex-valued activation function CReLU and the complex convolution operation (ComplexConv), where the kernel size is 3$\times$3 and the number of filters is $\mathbf{x}$ except for the last layer.} 
	\label{fig:CV-FCN}
\end{figure*}

The interference mitigation problem is tackled as a regression problem.
As targets' beat signals and interferences show distinct distributions in the $t$-$f$ domain, the $t$-$f$ domain spectral diagram obtained with the STFT algorithm is naturally used for IM. 
Considering the complex-valued nature of FMCW radar signals in the $t$-$f$ domain, a complex-valued fully convolutional network architecture is designed for interference mitigation with the basic complex-valued modules (see Fig.~\ref{fig:CV-FCN}). The proposed CV-FCN is composed entirely of complex convolution layers, each of which except the last convolution layer is followed by the complex-valued activation function CReLU. 
The number of filters is fixed to one in the last convolution layer, which is used only to produce the output.

The $t$-$f$ spectrum of the interfered radar signal is set as the input of the network, and its counterpart of the associated reference (i.e., the clean signal) is used as the label. 
Since the existing deep learning tools do not support the complex-valued input data, the real and imaginary parts of input samples are separated into two channels.
Meanwhile, the square kernels with size 3$\times$3 are used to deal with the two-dimensional input samples, and the zero-padding is used in the complex convolutional layer to ensure the output $t$-$f$ spectrum have the same shape as the input.

\begin{figure*}[!t]
	\centering
	\vspace{-3mm}
	\includegraphics[width=0.95\textwidth]{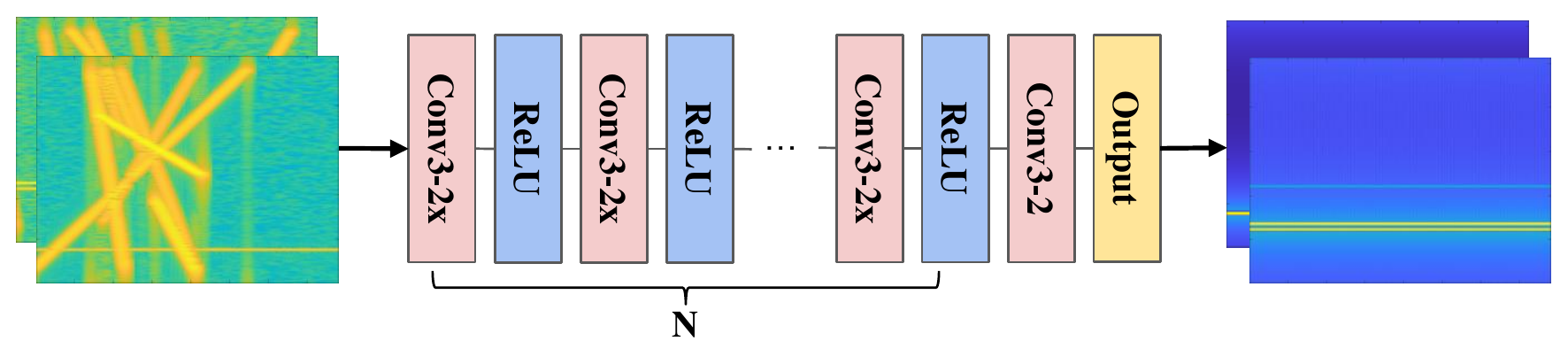}
	\caption{Proposed real-valued fully convolutional network architecture (RV-FCN) for comparison. It uses traditional real-valued ReLU and the convolution operation (Conv), where the kernel size is 3$\times$3 and the number of filters is $\mathbf{2x}$ except for the last layer.} 
	\label{fig:RV-FCN}
\end{figure*}

\subsection{Loss function}

MSE is generally used as a loss function in the DL-based interference mitigation approaches \cite{mse}, which calculates the difference between the output of the network and the related label. However, the performance of the network trained with the MSE as the loss function is limited by the size of the training dataset. With the increase of the network's total parameters and training iterations, the overfitting problem cannot be avoided, making it challenging to extract critical features. Moreover, due to the lack of real interfered radar signals and related reference data in practice, synthetic data based on the analytical signal models are commonly generated to build the training dataset. However, the analytical model used for data synthesis is generally derived based on certain assumptions (e.g., perfect radar system and frequency-independent scattering property of targets) for simplification, which may make the synthetic dataset impractical to contain all the features of the data acquired in various realistic scenarios. Consequently, the performance of the networks trained using the MSE with only simulated radar signals would degrade when utilized to real data. To avoid the overfitting problem and improve the generalization of the trained network, explicitly incorporating the prior information could be of benefit for real data. 

As shown in Section II, the interferences lead to time-varying beat frequencies after dechirping while the frequencies of targets' beat signals are constant. The projection of the interference on the frequency axis is a line, while the projections of targets' beat signals on the frequency axis are some points. Thus, interferences and targets' beat signals show different sparsities along the frequency axis (i.e., different sparsities in the range profiles). To exploit this prior feature of interferences and beat signals in the $t$-$f$ domain, we introduce the $L_{2,1}$ norm of the recovered $t$-$f$ domain spectrum (i.e., the output of the proposed neural network) as a regularization term for the NN training in addition to the traditional loss function MSE. The complete loss function is expressed as 
\begin{align}
	l(\mathbf{S},\Tilde{\mathbf{S}}) &= \|\mathbf{S} - \Tilde{\mathbf{S}}\|_F^2 + \lambda \|\Tilde{\mathbf{S}}\|_{2,1} \label{fmcw_loss} \\ 
	\|\Tilde{\mathbf{S}}\|_{2,1} &= \sum_{j=1}^N\sqrt{\sum_{i=1}^M|\Tilde{S}_{ij}|^2}\  \label{L21norm}
\end{align}
where $\|\mathbf{X}\|_F$ and $\|\mathbf{X}\|_{2,1}$ represent Frobenius norm and $L_{2,1}$ norm of a matrix $\mathbf{X}$, respectively, $\Tilde{\mathbf{S}} \in \mathbb{C}^{M\times N}$ is the matrix of the recovered spectrum in the $t$-$f$ domain with the row and column related to the frequency and time dimensions, respectively, and $\mathbf{S} \in  \mathbb{C}^{M\times N}$ is the label. $\|\mathbf{S} - \Tilde{\mathbf{S}}\|_F^2$ is the MSE loss function, and $\|\Tilde{\mathbf{S}}\|_{2,1}$ is the $L_{2,1}$ norm of $\Tilde{\mathbf{S}}$, as shown in \eqref{L21norm}. $\lambda$ is a hyper-parameter used to make a trade-off between the MSE (i.e., data consistency) and the prior feature.

Due to the introduced regularization term $\|\Tilde{\mathbf{S}}\|_{2,1}$, the overfitting problem can be avoided as much as possible. 
Moreover, as the regularization term $\|\Tilde{\mathbf{S}}\|_{2,1}$ provides solid expert knowledge, it boosts the convergence rate of the network training (i.e., the network can be trained with fewer iterations) and the size of the dataset needed for training can be significantly reduced. 
Besides, the feature used for IM is the fusion of the prior information and the features extracted from the CNNs, which can motivate better generalization capability.

\subsection{Training Setup}
Before being fed into the network, the complex-valued input samples are normalized. Specifically, the normalization method can be described as:
\begin{equation}
	\Tilde{\mathbf{Y}}(m,n) = \frac{\mathbf{Y}(m,n)}{\max\limits_{\substack{1\le m \leq M \\ 1\leq n\leq N} } \left|\mathbf{Y}(m,n)\right|_2 }
	\label{equa_data_norm}
\end{equation}
where $\mathbf{Y}$ is the matrix of the STFT spectrum of beat signals contaminated by interferences, and $m$ and $n$ are the row and column indices of an entry of the matrix.

After being processed by the network, the matrix of the recovered spectrum is multiplied by the denominator in equation (\ref{equa_data_norm}) for further processing.

In the network training process, the complex weight initialization strategy\cite{complex} was used to initialize the parameters of complex convolution layers. The Adam algorithm with a fixed learning rate of 0.001 and 32 input samples per batch was used for training. We end the training at epoch 100 after good convergence was observed. Moreover, all the models were trained on a single NVIDIA 2080Ti graphics processing unit (GPU). The code was implemented using Keras and Tensorflow tools.

\section{Datasets}

In this section, the setups of numerical simulations for data synthesis and experimental measurements are introduced in detail, and then the data split algorithm for a more flexible model is described.

\subsection{Radar signals synthesis}

\begin{table}[!t]
	\caption{Parameters of the victim radar}
	\centering
	\label{tab:parameter_victimRadar}
	\begin{tabular}{@{}llll@{}}
		\toprule
		\textbf{Parameter}                     & \textbf{Value}      & \textbf{Parameter}        & \textbf{Value}\\ 
		\midrule
		Center frequency                       & $3\,\mathrm{GHz}$   & Velocity                & $30\,\mathrm{km/h}$ \\
		Duration of a sweep $T_{\text{sw}}$    & $400\,\mu s$        & Window type             & Hamming  \\
		Bandwidth                              & $40\,\mathrm{MHz}$  & Window length           & $256$    \\
		Chirp rate $K$                         & $10^{11}\,\mathrm{Hz/s}$           & Overlap length          & $255$    \\
		Sampling frequency                     & $12\,\mathrm{MHz}$  & FFT points              & $256$    \\
		Maximum detection distance             & $8\,\mathrm{km}$    \\ 
		\bottomrule
	\end{tabular}
\end{table}

Due to the difficulties in acquiring both interfered radar echoes and their related references in practice, especially for dynamic scenarios, in this paper we decided to use synthetic FMCW radar signals for the proposed neural network training and then employ both synthetic and measured data for test.

For data generation, a victim FMCW radar with the parameters described in Table~\ref{tab:parameter_victimRadar} was considered.
To emulate the scenarios with various scatterers and different interferences, 
each parameter of targets and interfering signals was randomly chosen from a uniform distribution $\mathcal{U}[a,b]$ for continuous variables in a closed interval or $\mathcal{U}(a,b)$ for continuous variables in an open interval or $\mathcal{U}\{a,b\}$ for discrete variables, where $a$ and $b$ define the bounds of an interval. The detailed intervals of the values of the parameters of targets and interfering signals are shown in Table \ref{tab:parameters_targets_interfer}, where $K$ and $T_{sw}$ refer to the chirp rate and sweep duration of the victim radar in Table \ref{tab:parameter_victimRadar}. 
Moreover, complex white Gaussian noise was added to synthetic signals to account for system noise and measurement errors. To characterize the interference-contaminated signals in the presence of complex white Gaussian noise, Signal to Noise Ratio (SNR) and Signal to Interference plus Noise Ratio (SINR) are used as metrics. The SNR ranges from -20\,dB to 20\,dB with step size of 5\,dB while the SINR takes values randomly from a uniform distribution as shown in Table \ref{tab:parameters_targets_interfer}.

\begin{table}[!t]
	\caption{Parameters of the targets and interference}
	\centering
	\label{tab:parameters_targets_interfer}
	\begin{tabular}{@{}llll@{}}
		\toprule
		\textbf{\begin{tabular}[c]{@{}l@{}}Parameter \\of Targets\end{tabular}}      & \textbf{Value}       & \textbf{\begin{tabular}[c]{@{}l@{}}Parameter of \\Interferences\end{tabular}}   &\textbf{Value}        \\ 
		\midrule
		Number                  & $\mathcal{U}\{0, 20\}$              & Number               & $\mathcal{U}\{1,20\}$   \\ 
		Distance                & $\mathcal{U}(8, 8000)\,\mathrm{m}$  & Amplitude            & $\mathcal{U}(0,3)$      \\ 
		Amplitude               & $\mathcal{U}(0, 3)$                 & Center frequency     & $3\,\mathrm{GHz}$       \\ 
		Phase                   & $\mathcal{U}(0, 2\pi)$              & Chirp rate           & $\mathcal{U}(-2 K,2K)$  \\ 
		Velocity                & $\mathcal{U}(0, 80)\,\mathrm{km/h}$ & Duration             & $\mathcal{U}(0,T_{\text{sw}})$          \\ 
		SNR                     & $\mathcal{U}\{-20,20\}\,\mathrm{dB}$             &	Delay time           & $\mathcal{U}\left(\frac{-T_{\text{sw}}}{2},\frac{T_{\text{sw}}}{2}\right)$  \\ 
		SINR                    & $\mathcal{U}[-40,20]\,\mathrm{dB}$  \\
		\bottomrule
	\end{tabular}
\end{table}

\begin{figure}[!t]
	\centering
	\subfloat[Interfered signal]{\label{fig:simulated_interfered_signal}
		\includegraphics[width=0.25\textwidth]{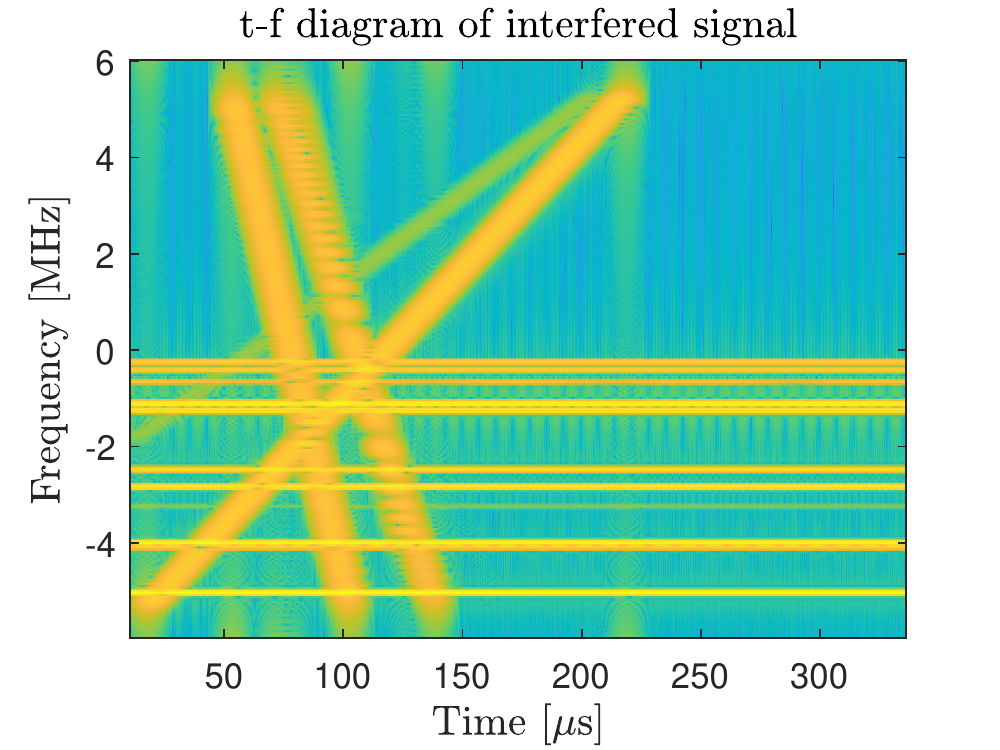}}
	\hspace{-5mm}
	\subfloat[Clean signal]{\label{fig:simulated_clean_signal}
		\includegraphics[width=0.25\textwidth]{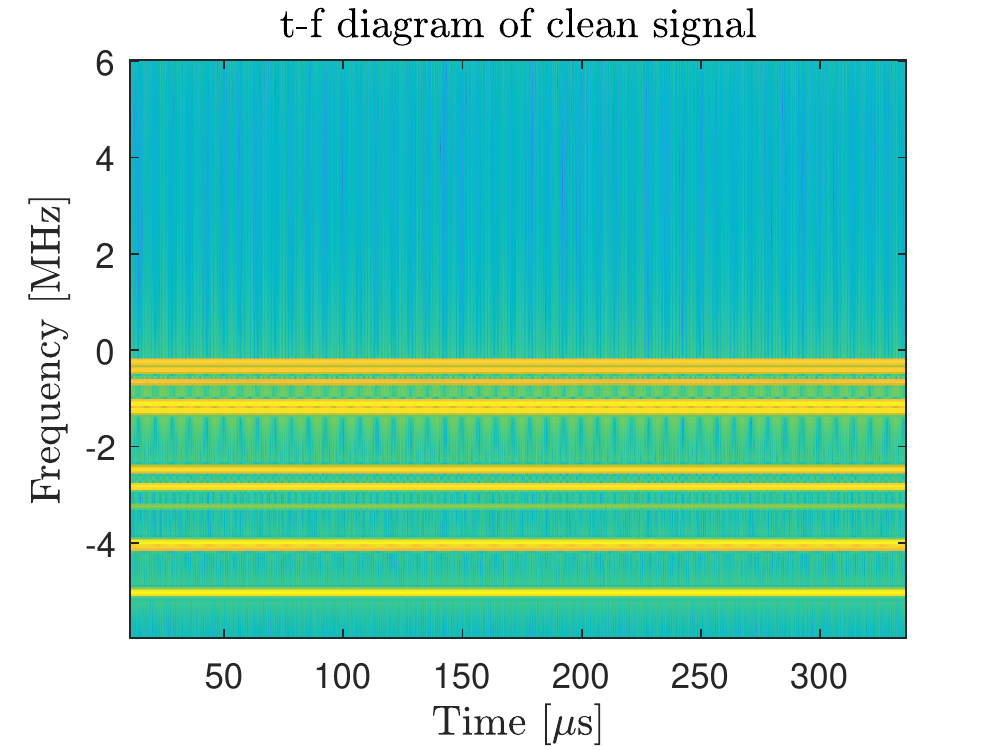}}
	\caption{$t$-$f$ diagram of simulated radar signals.}
	\label{fig:simulated_signal}
\end{figure}

After synthesizing the time-domain radar signals according to the setups described above, their time-frequency spectra are generated through the STFT algorithm. Specifically, the STFT was implemented by using a 256-point hamming window with a hop size of one for signal segmentation and 256-point fast Fourier transform (FFT) for spectrum calculation (see Table \ref{tab:parameter_victimRadar}). Since the beat signals of targets and interferences are synthesized according to controllable parameters, we can obtain both the interfered signals and the associated references (see the example in Fig~\ref{fig:simulated_signal}). 

\subsection{Experimental Measurements}

The experimental data in this paper were collected with the full-polarimetric PARSAX radar in TU Delft, which has two orthogonally polarized transmitting channels and four receiving channels for full polarimetric signal acquisition. 
We simultaneously use a horizontally polarized (H-pol) channel to emit a fixed FMCW signal and the vertically polarized (V-pol) channel to transmit an arbitrary FMCW-type waveform with various chirp rates, time duration, bandwidth, and time delay relative to the beginning of the signal in the H-pol channel. 

\begin{table}[!t]
	\caption{Parameters of the PARSAX Radar}
	\centering
	\label{tab:parameters_parsax}
	\begin{tabular}{ll}
		\toprule
		\textbf{Parameter}                 & \textbf{Value}       \\ 
		\midrule
		Center frequency                   & $3.315\,\mathrm{GHz}$  \\ 
		Duration of a sweep $T_{\text{sw}}$                & $1\,\mathrm{ms}$     \\ 
		Bandwidth                          & $30\,\mathrm{MHz}$   \\
		Chirp rate $K$                     & $30\,\mathrm{MHz/ms}$                   \\
		Sampling frequency                 & $400\,\mathrm{MHz}$  \\
		\bottomrule
	\end{tabular}
\end{table}

\begin{figure}[!t]
	\centering
	\subfloat[Camera visual image]{
		\label{fig:measured_camera}
		\includegraphics[width=0.25\textwidth]{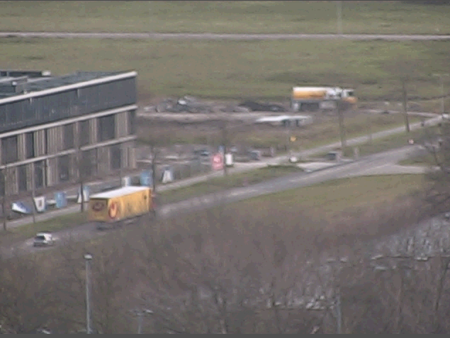}}
	\subfloat[$t$-$f$ diagram]{
		\label{fig:measured_t_f_diagram}
		\includegraphics[width=0.25\textwidth]{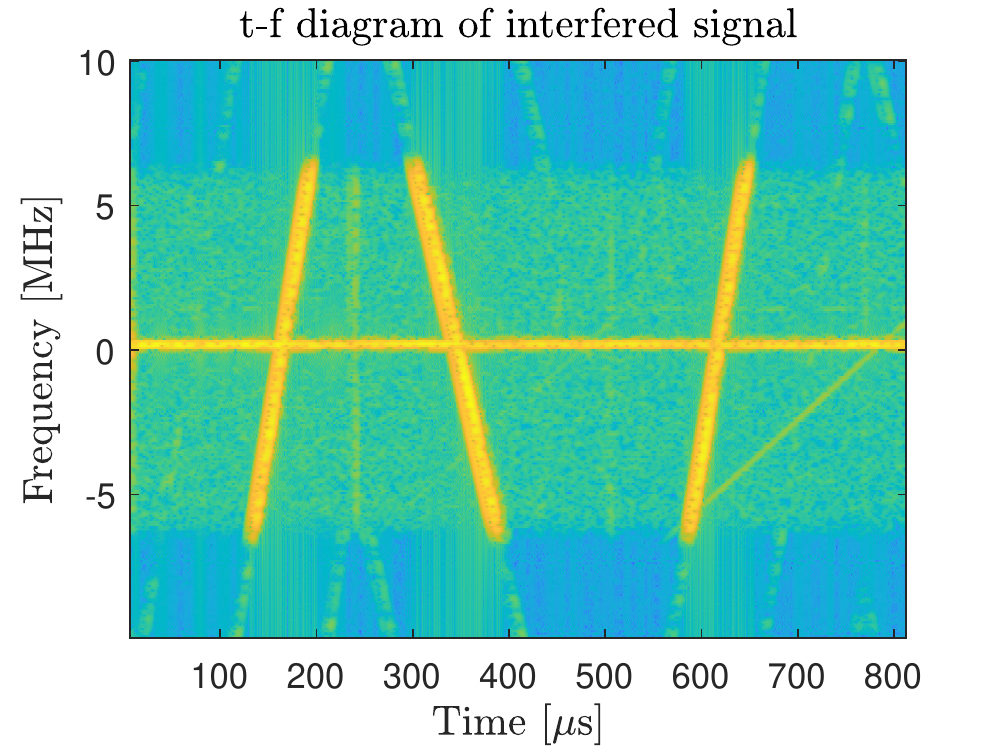}}
	\caption{Measured radar signals collected in the street scenario.}
	\label{fig:measured_radar_signal}
\end{figure}

The full-polarimetric signals scattered from the illuminated scene arrive at the receiving antenna at the same time. 
After passing through an orthomode transducer, the H-pol (i.e., HH and VH) and V-pol (i.e., HV and VV) scattered signals can be separated. 
However, the HH and VH (correspondingly HV and VV) signals inevitably interfere with each other at the receiving channels. 
As the HH (correspondingly VV) signals are generally much stronger than the VH (correspondingly HV) signals, the interference impact of HH (VV) on the VH (HV) is generally much severer.  
So, the acquired HV signals are used to construct the experimental dataset used in this paper.
The radar data were measured by illuminating three scenes: an industrical chimney, a rotating wind turbine, and a street with moving cars.
In total, 500-sweep radar data were measured with various interference signals for each sweep. As the latter two scenes were dynamic, acquiring the related references with our radar system was impractical, which is generally the case in practice. As an example, Fig. \ref{fig:measured_radar_signal} illustrates the street scenario at a time instant and the $t$-$f$ spectrum of the acquired signal. As the references are unavailable, the experimental data are only used to test the trained neural networks.

\subsection{Data Split}

In principle, the shape of the STFT spectra of radar signals is determined by the number of sampling points in a sweep and parameters of the STFT algorithm, and most CNNs can only process input samples of specific shapes.
In order to process radar signals of different shapes in the $t$-$f$ domain, the matrix of the STFT spectrum is split into a combination of smaller matrices before being fed into the network, which can be described as Algorithm \ref{algo_disjdecomp}.
Each element in the matrix of the recovered spectrum is predicted according to both the past and future time-frame information, and the phase is guaranteed to be continuous. 
In our experiments, the $N_p$ is set to 4, and the shape of the input sample (i.e., $M\times M$) is 256$\times$256.

\IncMargin{1em}
\begin{algorithm}[!t]
	\SetKwData{Left}{left}\SetKwData{This}{this}\SetKwData{Up}{up}
	\SetKwFunction{abs}{abs}\SetKwFunction{ModelPredict}{ModelPredict}\SetKwFunction{Max}{Max}
	\SetKwInOut{Input}{input}\SetKwInOut{Output}{output}
	
	\Input{Interfered STFT maps $\mathbf{Y}$ ($L\times N\times M$, $L$ number of maps, $N$ time samples per chirp, $M$ number of FFT points). $N_p$, number of overlap points}
	\Output{Recovered STFT maps $\Tilde{\mathbf{S}}$ ($L\times N\times M$)}
	\BlankLine
	\tcp{Data split}
	$p$ = $N/(M-2N_p) + 1$\;
	$\mathbf{T}[:L,:,:]$$=$$\mathbf{Y}[:,:M,:]$\;
	\For{$i= 1$ \KwTo $(p-2)$}{
			$\mathbf{T}[i\times L:(i+1)\times L,:,:]$$=$$\mathbf{Y}[:,i\times(M-2N_p):i\times(M-2Np)+M,:]$\;
		}
 	$\mathbf{T}[(p-1)\times L:p\times L,0:N-(p-1)\times(M-2N_p),:]$ = $\mathbf{Y}[:,(p-1)\times(M-2N_p):N,:]$\;
 	\tcp{Data normalization}
 	\For{$i = 0$ \KwTo ($p\times L$)}{
 		$scl[i]$ $=$ \Max(\abs($\mathbf{T}[i,:,:]$))\;
 		$\mathbf{T}[i,:,:]$$=$$\mathbf{T}[i,:,:]$ / $scl[i]$)\;
 		}
 	$\Tilde{\mathbf{T}}$ = \ModelPredict{$\mathbf{T}$}\;
 	\tcp{Data denormalization}
 	\For{$i = 0$ \KwTo ($p\times L$)}{
		$\Tilde{\mathbf{T}}[i,:,:]$$=$$\Tilde{\mathbf{T}}[i,:,:]$ $\times$ $scl[i]$)\;
		}
	\tcp{Data integration}
	$\Tilde{\mathbf{S}}[:,:M-N_p,:]$ = $\Tilde{\mathbf{T}}[:L,:M-N_p,:]$\;
	\For{$i = 1$ \KwTo $(p-2)$}{
	$\Tilde{\mathbf{S}}[:,i\times(M-2N_p)+N_p:i\times(M-2N_p)+M-N_p,:]$ = $\Tilde{\mathbf{T}}[i\times L:(i+1)\times L,N_p:M-N_p,:]$\;
	}
	$\Tilde{\mathbf{S}}[:,(p-1)\times(M-2N_p)+N_p:N,:]$ = $\Tilde{\mathbf{T}}[(p-1)\times L:p\times L,N_p:N-(p-1)\times(M-2N_p),:]$\;
	
	\caption{Data processing}\label{algo_disjdecomp}
\end{algorithm}\DecMargin{1em}

\section{Simulation Results} \label{sec:simu_results}
In this section, the prior-guided deep interference mitigation approach based on the CV-FCN proposed in Section III-B is analyzed and demonstrated using synthetic FMCW radar signals. 
Firstly, the performance metric used for the quantitative evaluation of interference mitigation performance in our experiments is presented. 
Then the optimal network architecture based on the CV-FCN is obtained by grid search using MSE as the loss function, including the size and number of filters in each convolution layer, depth of the network, and additional residual connection. 
Next, the CV-FCNs are compared with the real-valued counterparts over a variety of network depths to show the superiority of complex-valued representation. After that, the prior-guided loss function is used for training, and its effects on the training iterations (i.e., convergence rate) and the size of the training dataset are investigated as well. Finally, our proposed approach is compared with the state-of-the-art conventional interference mitigation algorithms.

\subsection{Performance Metrics}
To quantitatively evaluate the performance of different interference mitigation methods, the SINR of a recovered radar signal relative to the clean reference is used as a performance metric. The SINR cannot only measure the remaining interferences and noise in the recovered signal, but also represent the signal distortion. It is defined as:
\begin{align}
	\text{SINR} = 10\lg \left(\frac{|\mathbf{s}|_2^2}
	{|\Tilde{\mathbf{s}}-\mathbf{s}|_2^2}\right)
\end{align}
where $\Tilde{\mathbf{s}}$ is the recovered signal in the time domain, and $\mathbf{s}$ is the corresponding reference.
Note the SINR is inversely proportional to the error vector magnitude (EVM) \cite{CNN}.

\subsection{Network architecture optimization}

\begin{figure}[!t]
	\centering
	\includegraphics[width=0.4\textwidth]{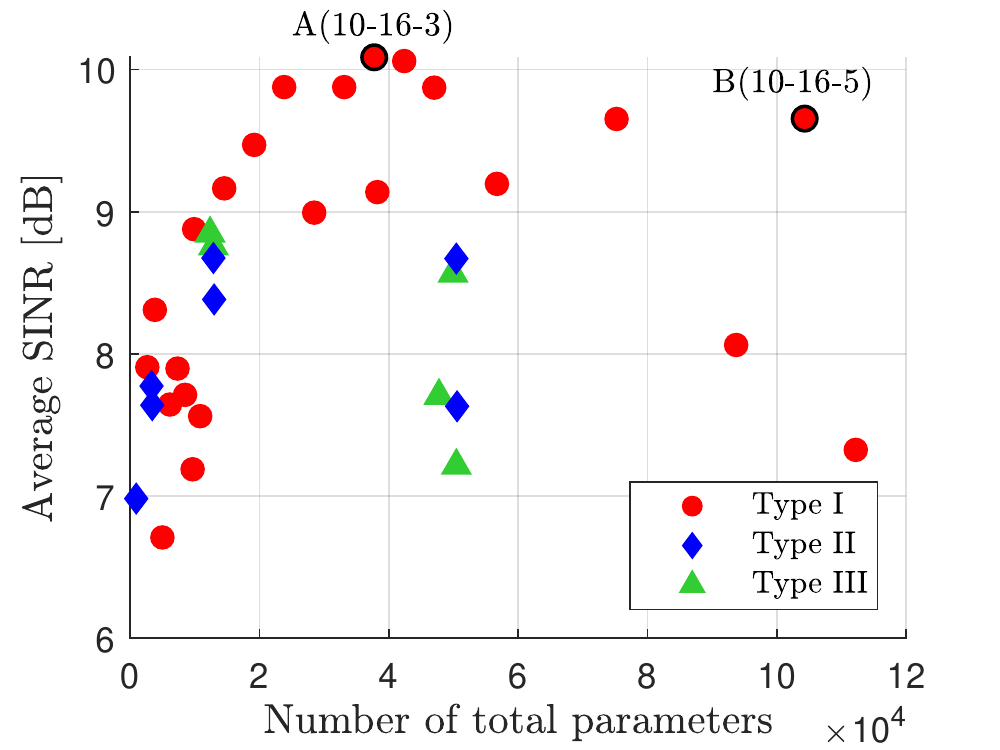}
	\caption{The performance comparison of different network architectures based on the CV-FCN trained using MSE as the loss function, where 10-16-m means the network has three convolution layers, the number of filters in each layer is 16 and the kernel size is m$\times$m.} 
	\label{fig:Grid_search_FCN}
\end{figure}

To find an optimal network architecture based on the proposed CV-FCN for FMCW radar interference mitigation, a parameter search is performed using MSE as the loss function.

Firstly, the problem of how to design the number of filters in each convolution layer is investigated.
Three different network architectures denoted as Type I, II, and III are implemented for comparison. Except for the last layer, the number of filters in each convolution layer is fixed to constant in Type I, and the number of filters is doubled or halved in each convolution layer for Type II and Type III, respectively.
The performance comparison of different architectures is shown in Fig. \ref{fig:Grid_search_FCN}, the average SINR of all evaluated architectures is given, and the x-axis indicates the number of total parameters of the network.
We change the number of parameters by varying the depth of the network, the number of filters in each layer and the kernel size.
We can see from the figure that the CV-FCNs where the number of filters is fixed to constant  (Type I) obtain better results. 
In fact, the number of filters is doubled in each convolution layer for the famous CNN architectures such as VGG \cite{simonyan2014very}.
Due to the use of the pooling layer in VGG, the number of channels is doubled in each layer to ensure the information amount of the connected convolution layers does not differ too much.
Since the pooling layer is not used in the CV-FCN to avoid signal distortion, the architecture where the number of filters is fixed to constant is more suitable.

Based on the conclusion, the CV-FCNs (Type I) with different numbers of total parameters are compared as shown in Fig. \ref{fig:Grid_search_FCN}. 
The features of targets' beat signals and interferences in the $t$-$f$ domain are relatively simple, a larger number of filters or a deeper network is unnecessary.
It follows that the maximum average SINR is obtained using a model (Model A) with ten layers and a kernel size of 3$\times$3. Except for the last layer, the number of filters in each convolution layer is fixed to 16. 
Moreover, we compared the CV-FCNs with kernels of 3$\times$3 (Model A) and 5$\times$5 (Model B). 
Although the number of total parameters in Model B triples, it leads to an average SINR 0.43\,dB lower than that by Model A. 
So, the kernel size of 3$\times$3 is used in the following experiments.

\begin{figure}[!t]
	\centering
	\includegraphics[scale=0.8]{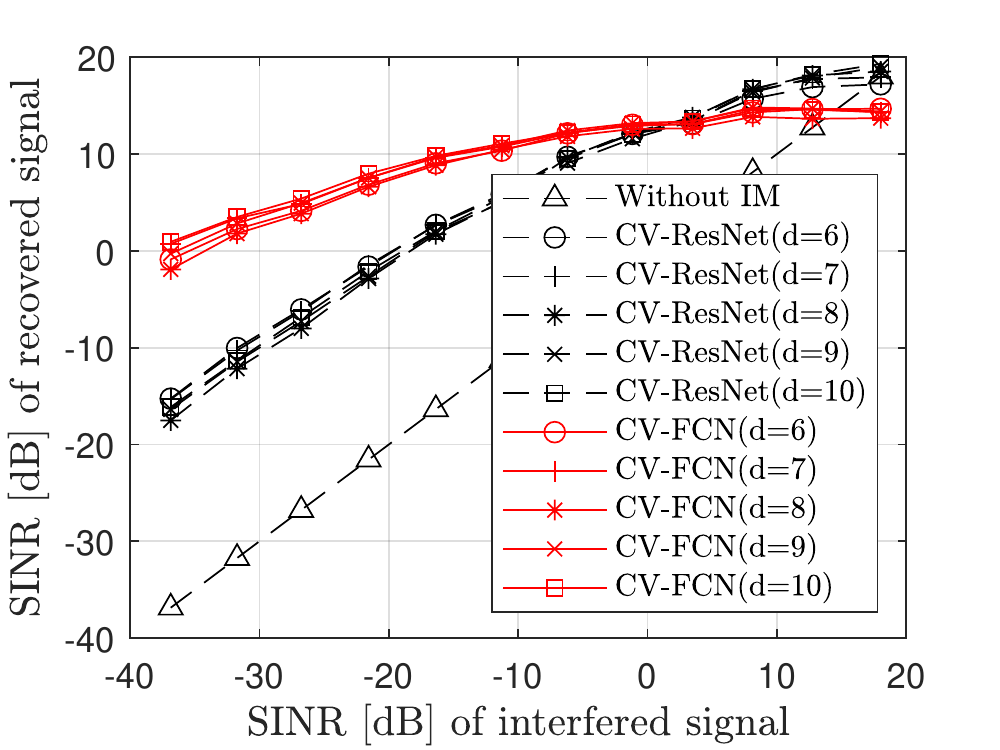}
	\caption{The performance comparison of the CV-FCN with CV-ResNet, where d means the number of convolution layers in the network.} \label{fig:residual connection}
\end{figure}

In recent years, researchers\cite{resnet} have shown that introducing the residual connection can promote better backpropagation of gradient, avoiding the problems of gradient diminishing and explosion during training. 
To analyze the effect of residual connection for complex-valued networks, an additional residual connection is added between the input and output of the CV-FCN, resulting in a complex-valued residual network (CV-ResNet) used for comparison with CV-FCN. 
Except for one additional residual connection, the other layers and parameters remain unchanged. 
The performance comparison is shown in Fig.~\ref{fig:residual connection}, and the result indicates that the CV-FCN has a better performance especially in low SINR scenarios. 
The beat frequency of the targets' signals is constant, which results in the horizontal lines in the $t$-$f$ spectrum. 
The features of the targets' beat signals can be extracted by the convolution filters of CV-FCN, while the filters in CV-ResNet deal with the features of interferences, which show inclined lines with different slopes, intensities, and durations in the $t$-$f$ spectrum and are relatively more complex.
Different features extracted explain why the residual connection does not work well.
However, the CV-ResNet may acquire better performance in other IM applications where signals are more complex than interferences.

\subsection{Performance comparison with real-valued networks}
In order to analyze the effect of complex-valued networks in the radar signal processing chain, we compared the CV-FCNs with their real-valued counterparts over a variety of network depths.

\begin{figure}[!tb]
	\centering
	\vspace{-4mm}
	\includegraphics[scale=0.8]{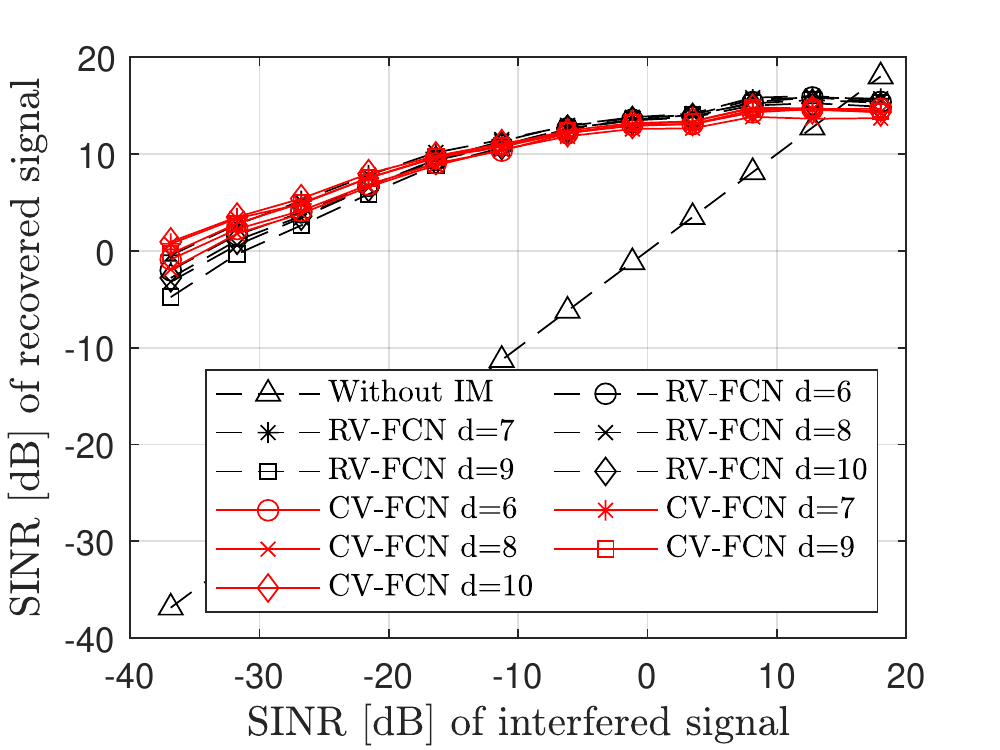}
	\caption{The performance comparison of the CV-FCNs with the corresponding RV-FCNs, where d means the number of convolution layers in the network.} \label{fig:complex}
\end{figure}

The real-valued FCN (RV-FCN) is constructed using the real-valued convolution layer and the ReLU activation function to replace the corresponding complex modules. In the complex-valued networks, the number of complex filters in each layer is the number of complex feature maps. It is also the effective number of feature maps for each of the real and imaginary parts. To obtain the same number of feature maps for performance comparison, the number of filters in each layer in RV-FCNs is twice that in CV-FCNs, and the other parameters remain unchanged (as shown in Fig. \ref{fig:RV-FCN}). 

The performance comparison is shown in Fig. \ref{fig:complex}, where the CV-FCNs of different depths show a better performance in low SINR scenarios compared to the real-valued counterparts. 
In the interference mitigation task, it is more meaningful to consider the performance in low SINR scenarios. 
The detailed parameters and quantitative results of the CV-FCNs and RV-FCNs are shown in Table \ref{tab:compared_to_real}. When the SINR of interfered signals is between -40\,dB and -20\,dB, all the CV-FCNs with different depths show better results (in bold red font), and the SINRs of the beat signals recovered by CV-FCNs are on average 1.1\,dB higher than that by RV-FCNs.
Moreover, due to complex multiplication and half the number of filters in each layer in the CV-FCN, its parameter amount is half of the real-valued counterparts. 

Next, we compared the optimal CV-FCN, whose total parameters are 37730, with the optimal RV-FCN whose total parameters are 84418. 
With the number of total parameters reduced by 55.3\%, the CV-FCN achieves almost the same performance as the RV-FCN. 
In the computer's memory, the optimal RV-FCN requires 1.04 megabytes of memory, while the CN-FCN requires only 525 kilobytes.
Additionally, the SINR of recovered signals has improved from -0.54\,dB to 0.93\,dB when the SINR of the interfered signals is between -40\,dB and -35\,dB.
This suggests that the CV-FCN can be better used in practical applications due to the limitation of the hardware memory and its superior performance in suppressing strong interferences.

\begin{table}[!t]
	\centering
	\begin{threeparttable}[b]
		\caption{Comparison of the CV-FCN with RV-FCN}
		\label{tab:compared_to_real}
		\centering
		\begin{tabular}{@{}ccccc@{}}
			\toprule
			\textbf{Method} &
			\textbf{\begin{tabular}[c]{@{}c@{}}filter \\ number\end{tabular}} &
			\textbf{depth} &
			\textbf{parameter} &
			\textbf{\begin{tabular}[c]{@{}c@{}}SINR (dB) in low SINR\\ scenarios\tnote{1}\end{tabular}} \\
			\midrule
			&                      & 6  & 19170 & {\color[HTML]{CB0000} \textbf{3.0872}} \\  
			&                      & 7  & 23810 & {\color[HTML]{CB0000} \textbf{4.1154}} \\  
			&                      & 8  & 28450 & {\color[HTML]{CB0000} \textbf{2.5828}} \\ 
			&                      & 9  & 33090 & {\color[HTML]{CB0000} \textbf{3.7518}} \\  
			&                      & 10 &
			{\color[HTML]{CB0000} \textbf{37730}} &
			{\color[HTML]{CB0000} \textbf{4.4528(optimal)}} \\
			&                      & 11 & 42370	& {\color[HTML]{CB0000} \textbf{4.1615}} \\
			\multirow{-7}{*}{\textbf{CV-FCN}} &
			\multirow{-7}{*}{16} 
			                       & 12 & 47010	& {\color[HTML]{CB0000} \textbf{4.0407}} \\ 
			\midrule
			&                      & 6  & 38178 & 2.5821                                 \\  
			&                      & 7  & 47426 & 3.8125                                 \\  
			&                      & 8  & 56674 & 1.8419                                 \\  
			&                      & 9  & 65922 & 0.8446                                 \\  
			&                      & 10 & 75170 & 2.1030                                \\  
			&                      & 11 & 84418 & 4.0085(optimal)                        \\
			\multirow{-7}{*}{\textbf{RV-FCN}} & 
			\multirow{-7}{*}{32} 
								   & 12 & 93666 & 2.8022                                 \\
			\bottomrule
		\end{tabular}
		\begin{tablenotes}
			\footnotesize
			\item[1] The SINR of interfered signals is between -40\,dB and -20\,dB.
		\end{tablenotes}
	\end{threeparttable}
\end{table}

\begin{figure}[!t]
	\centering
	\vspace{-4mm}
	\includegraphics[width=0.5\textwidth]{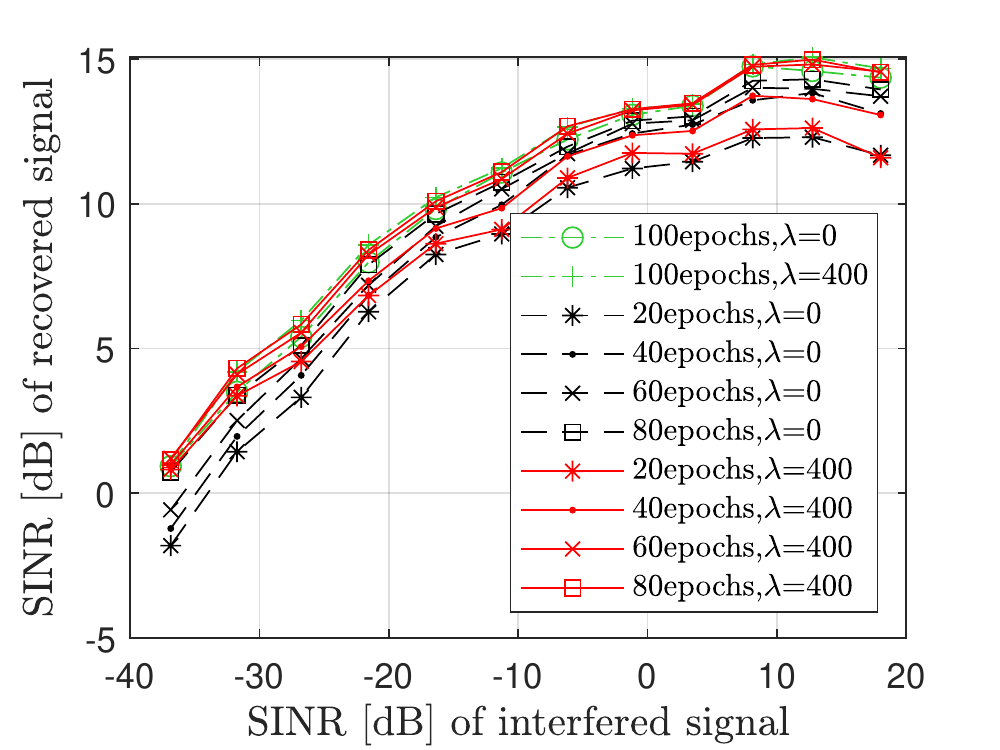}
	\caption{The performance comparison of CV-FCNs trained using all the 4320 samples.} \label{fig:data_reduction_all}
\end{figure}

\subsection{Effects of prior-guided loss function}\label{subsec:Effect_prior_LF}

In Section III-C, we proposed the prior-guided loss function based on the different distributions of targets' beat signals and interferences in the spectrogram. 
In this part, we use the prior-guided loss function instead of MSE to train the obtained optimal CV-FCN (Model A) in Section V-B. 
The hyper-parameter $\lambda$ in equation (\ref{fmcw_loss}) is used to make a trade-off between data consistency and prior knowledge. 
Note that the prior-guided loss function becomes the MSE when $\lambda=0$.

\subsubsection{Effect on convergence rate of training}\label{subsubsec:Effect_on_convergence} 
a training dataset containing 4320 samples ($t$-$f$ maps) is generated. Then, without loss of generality, $\lambda=0$ and $\lambda=400$ were used in the loss function \eqref{fmcw_loss} for comparison, and in each case the CV-FCN was trained for different epochs ranging from 20 to 100 with a step size of 20.
The performance of the obtained CV-FCNs is shown in Fig.~\ref{fig:data_reduction_all}. One can see that the performance of the CV-FCN improves with the increase of training epochs for both cases of $\lambda=0$ and $\lambda=400$. 
In low SINR scenarios, the CV-FCN trained for fewer epochs achieves comparable results as that trained for 100 epochs when the prior information was incorporated (i.e., $\lambda=400$).

\begin{figure*}[!t]
	\centering
	\vspace{-5mm}
	\subfloat[]{\label{fig:dataset_half}
		\includegraphics[width=0.3\textwidth]{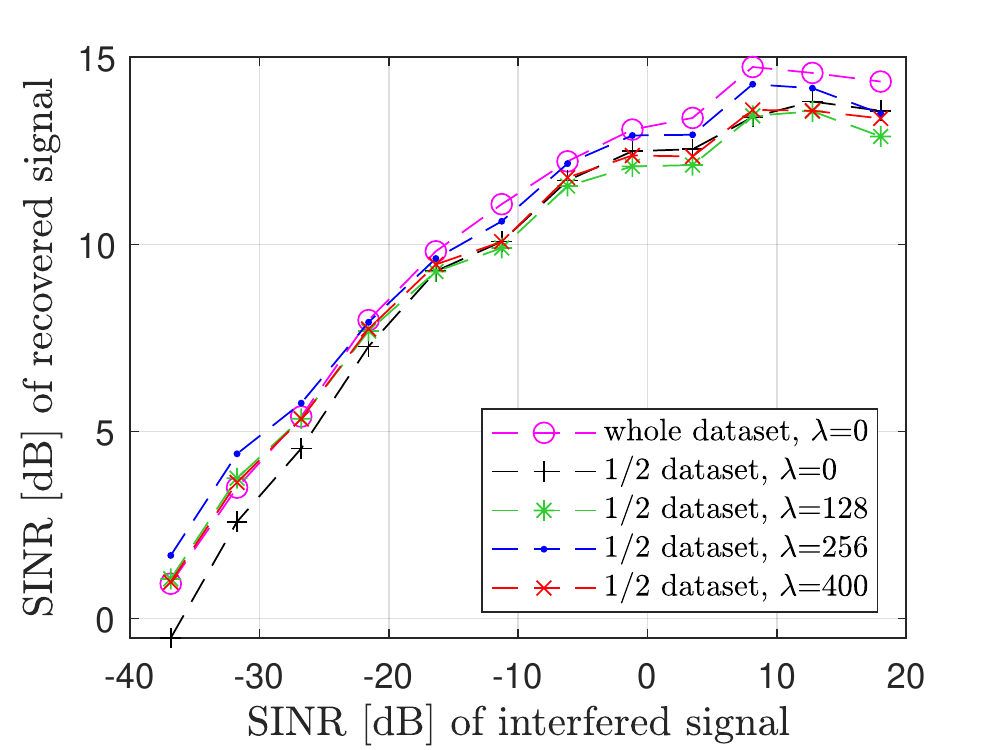}}
	\subfloat[]{\label{fig:dataset_quarter}
		\includegraphics[width=0.3\textwidth]{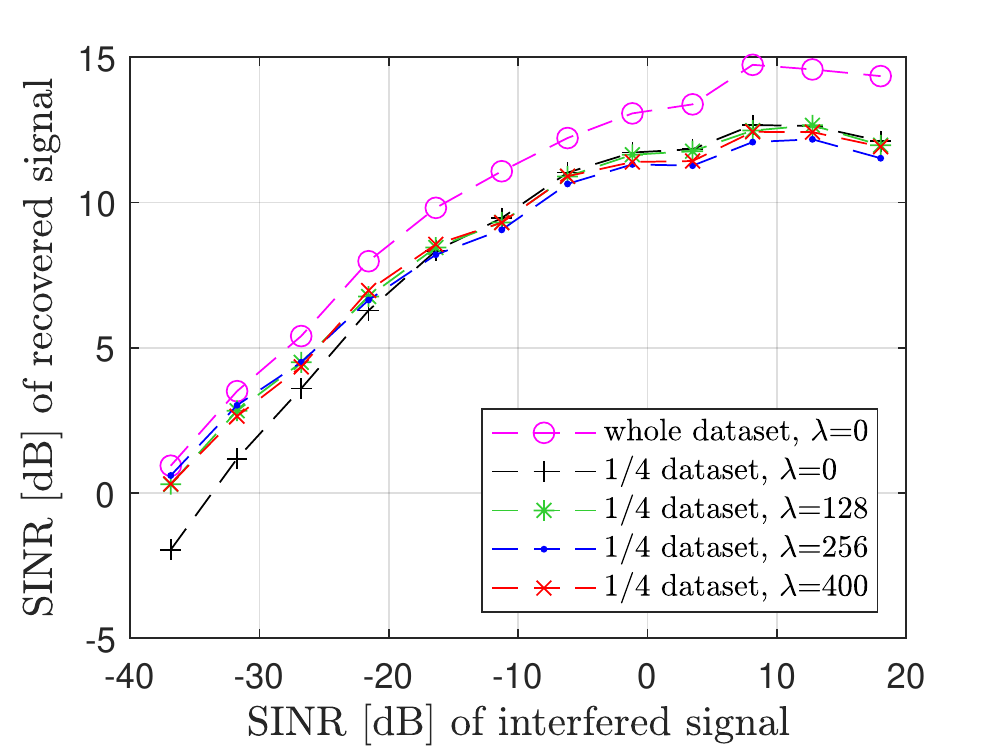}}
	\subfloat[]{\label{fig:dataset_eighth}
		\includegraphics[width=0.3\textwidth]{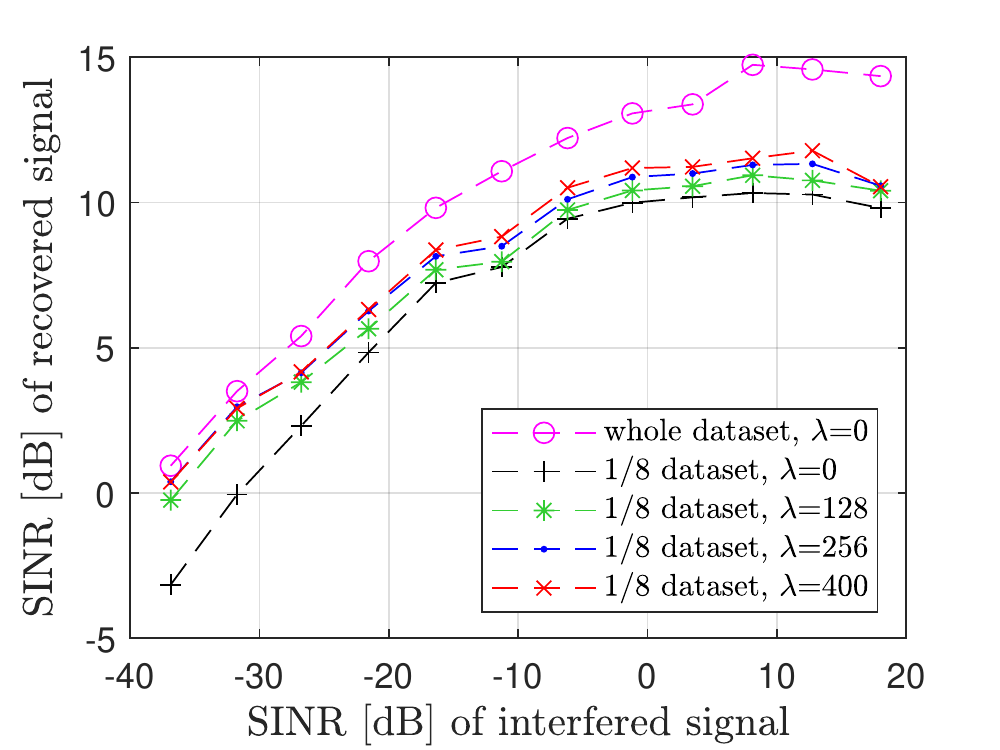}}
	
	\caption{Performance comparison of CV-FCNs trained with the datasets of different sizes using the prior-guided loss function. \protect\subref{fig:dataset_half} 1/2 dataset (2160 samples), \protect\subref{fig:dataset_quarter} 1/4 dataset (1080 samples) and \protect\subref{fig:dataset_eighth} 1/8 dataset (540 samples).}
	\label{fig:reduce_dataset}
\end{figure*}

\begin{figure}[!t]
	\centering
	\vspace{-4mm}
	\includegraphics[width=0.5\textwidth]{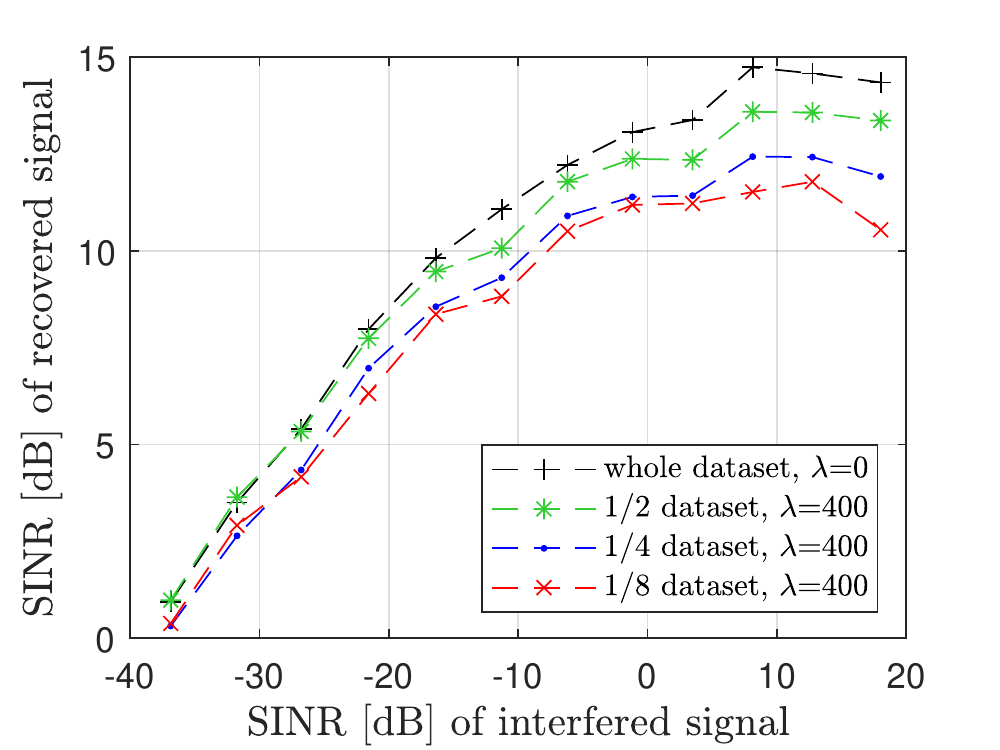}
	\caption{The performance comparison of CV-FCNs trained using different size of the dataset.} \label{fig:data_reduction}
\end{figure}

The network has not yet learned enough features to suppress the interference components as the training epochs are less than 100. 
With the introduced regularization term, the features needed for interference mitigation can be extracted faster by training.
On the other hand, the CV-FCN using the MSE as the loss function (i.e., $\lambda=0$) converged when trained for 100 epochs according to Fig.~\ref{fig:data_reduction_all}.
The MSE is sufficient to help the network to extract the correct features, and introducing $L_{2,1}$ norm does not make a difference in the performance.
As a result, the CV-FCN trained using the prior-guided loss function where $\lambda$ is 400 reaches almost the same results in low SINR scenarios with only 20 training epochs.
By contrast, in the high SINR scenarios, the interference components are reduced, and the noise becomes the dominant disturbance to the signal. 
Thus the MSE becomes the key part in the prior-guided loss function. 
Moreover, the regularization impact of $L_{2,1}$ norm becomes weaker for a fixed value of $\lambda$. 
This can explain the fact that the performance improvement in high SINR scenarios is not obvious.
One possible solution is to adjust the hyper-parameter $\lambda$ according to the SNR to overcome this problem.

\subsubsection{Effect on the size of training dataset}
To evaluate the effect of prior knowledge incorporation in the loss function on the size of the dataset required for training, three datasets of different sizes, i.e., $1/2$, $1/4$, and $1/8$ of the training dataset in section~\ref{subsubsec:Effect_on_convergence}, were generated. Based on the convergence analysis in section~\ref{subsubsec:Effect_on_convergence}, we set the training epochs to 100. 
Fig. \ref{fig:reduce_dataset} shows the performance of the CV-FCNs trained using the datasets of different sizes when $\lambda$ takes various values. According to Fig.~\ref{fig:reduce_dataset}\subref{fig:dataset_half}-\subref{fig:dataset_eighth}, the SINRs of recovered signals generally degrade with the decrease of the sizes of training datasets.  
The smaller size of the training dataset used, the severer the SINR degradation compared to that obtained with the full dataset. However, with the increase of $\lambda$, the SINRs of recovered signals have been improved especially in low SINR scenarios, and a more noticeable improvement can be seen for a smaller training dataset. 

\begin{figure*}[!t]
	\centering
	\vspace{-5mm}
	\subfloat[]{\label{fig:simu_sig_inp_RealPart}
		\includegraphics[width=0.3\textwidth]{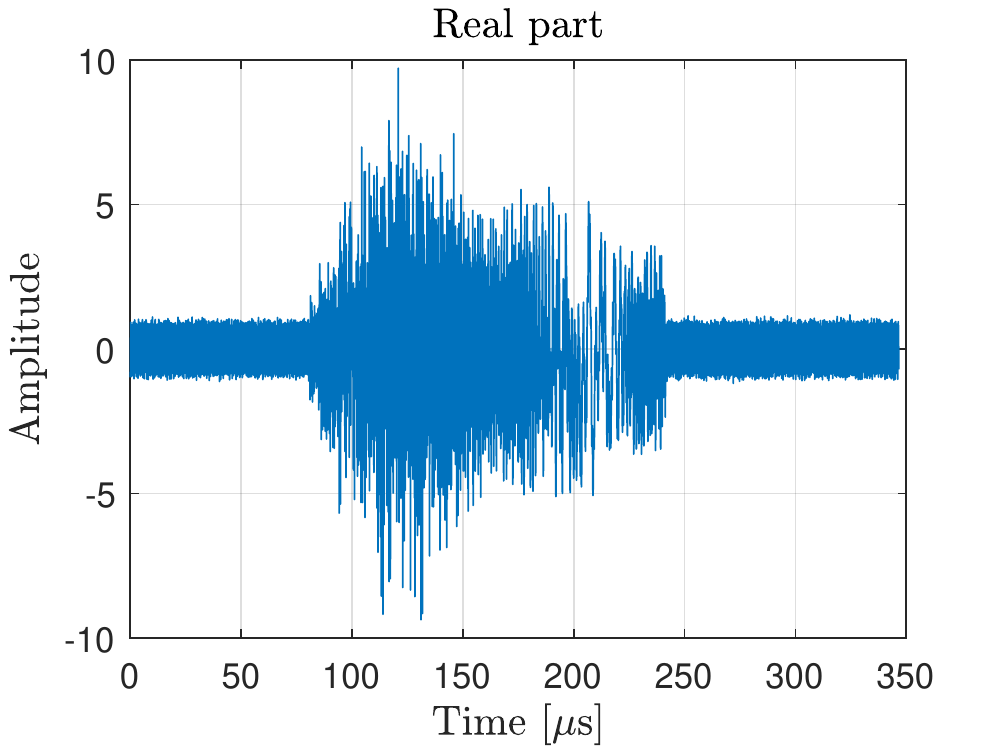}}
	\subfloat[]{\label{fig:simu_sig_fft_oral_inp}
		\includegraphics[width=0.3\textwidth]{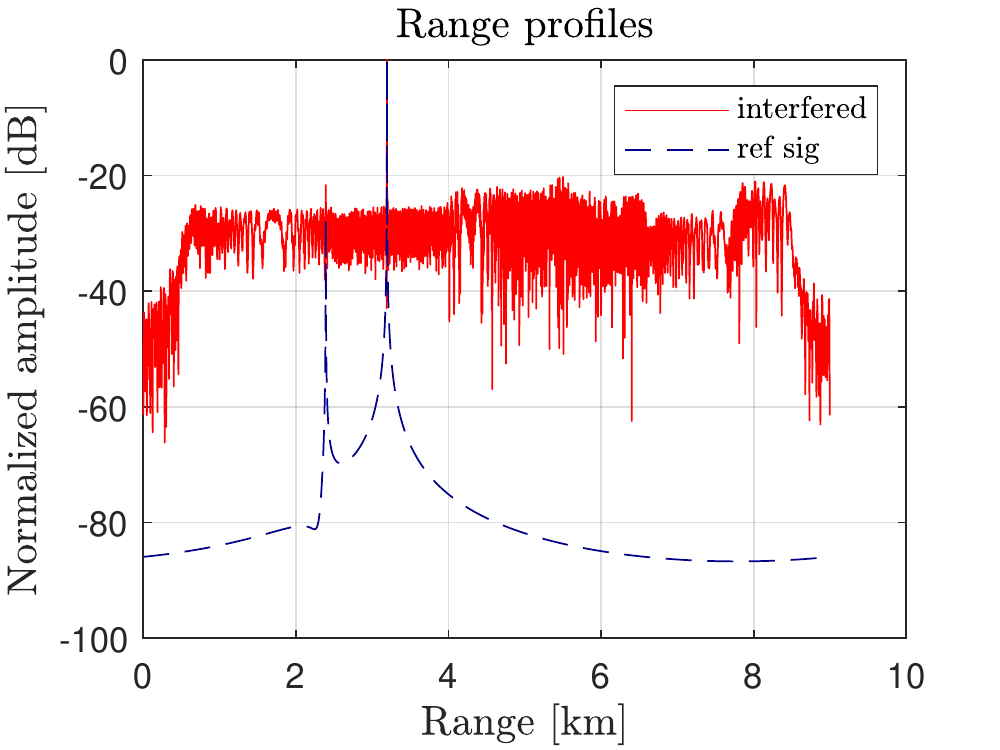}}
	\subfloat[]{\label{fig:simu_sig_TF_inp}
		\includegraphics[width=0.3\textwidth]{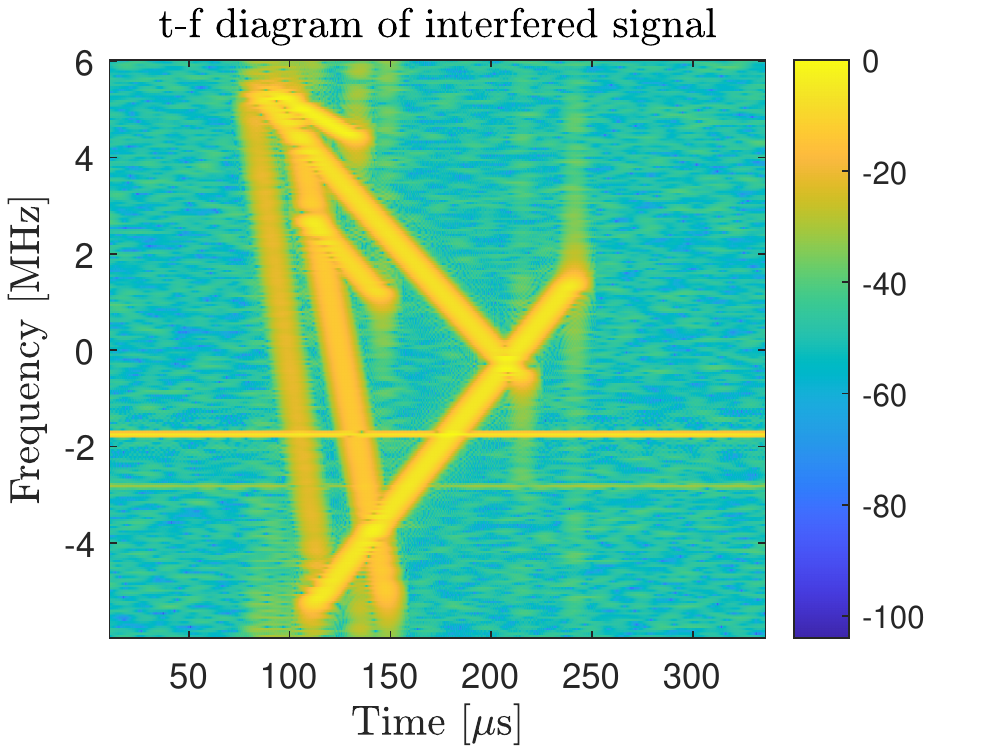}}
	
	\vspace{-3mm}
	\subfloat[]{\label{fig:simu_sig_TF_oral}
		\includegraphics[width=0.3\textwidth]{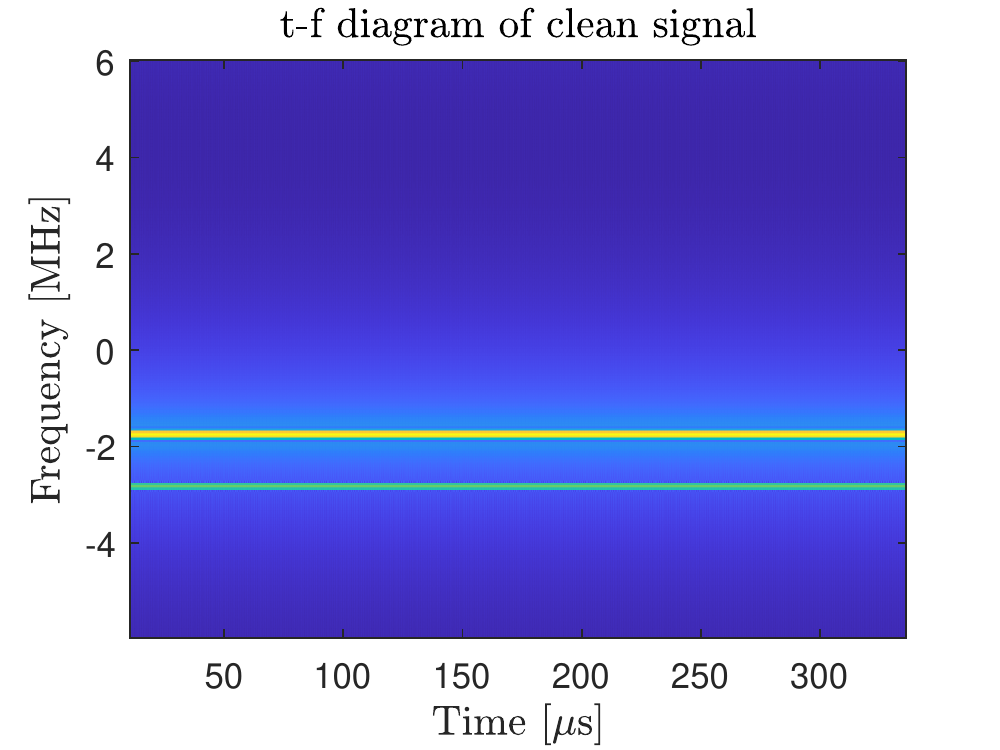}}
	\subfloat[]{\label{fig:simu_sig_TF_pred_b0}
		\includegraphics[width=0.3\textwidth]{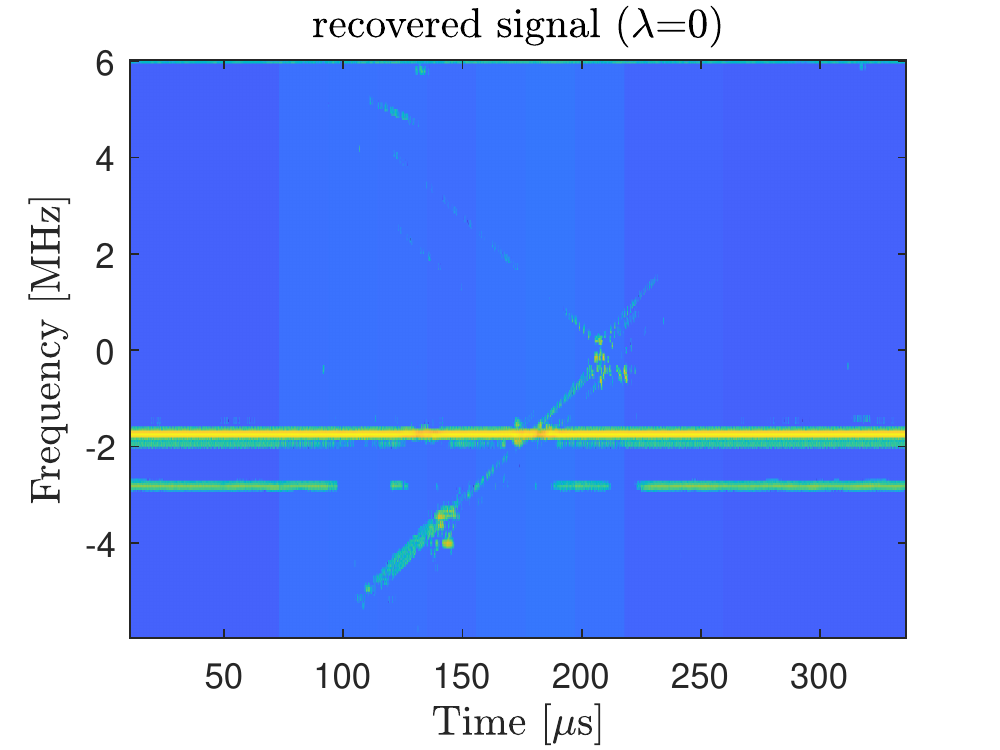}}
	\subfloat[]{\label{fig:simu_sig_TF_pred_b128}
		\includegraphics[width=0.3\textwidth]{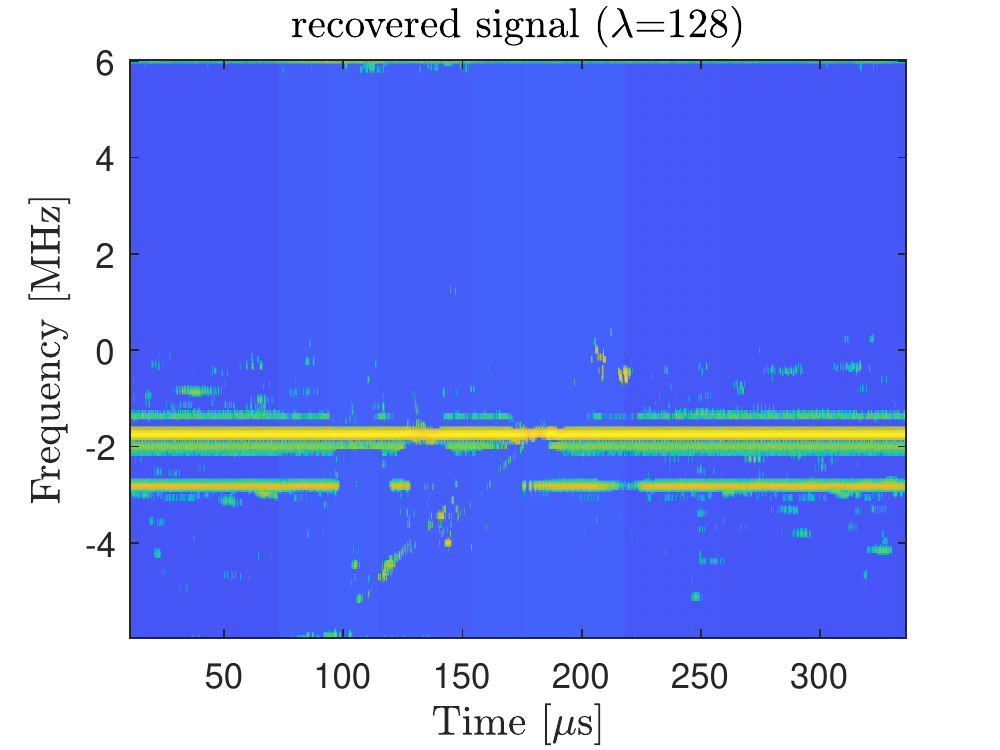}}
	
	\vspace{-3mm}
	\subfloat[]{\label{fig:simu_sig_TF_pred_b256}
		\includegraphics[width=0.3\textwidth]{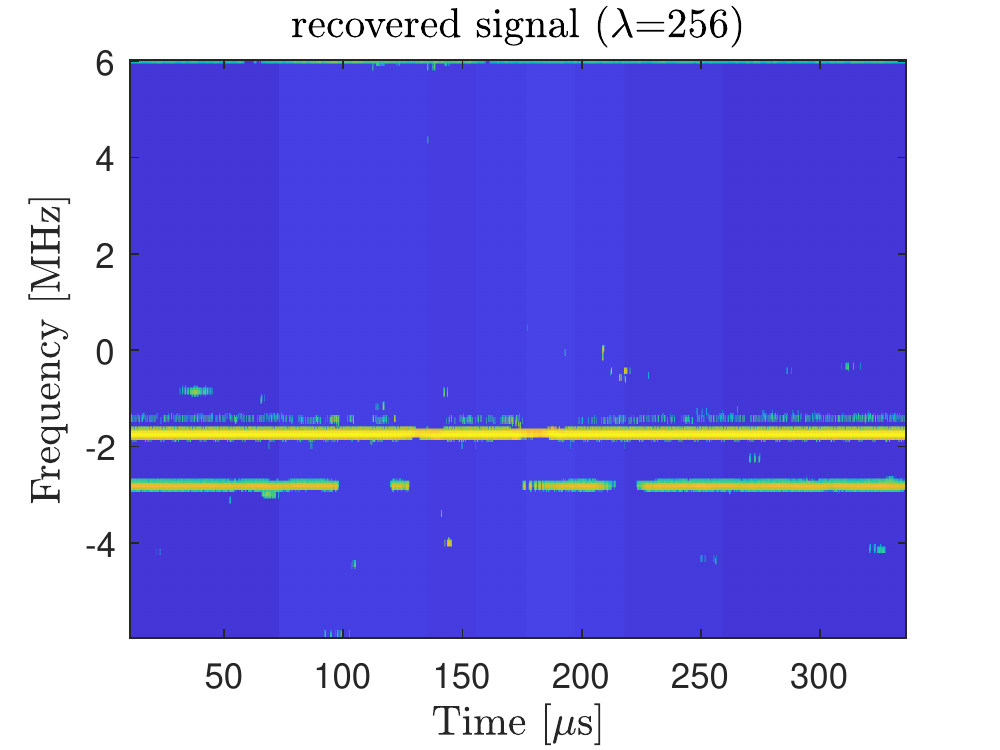}}
	\subfloat[]{\label{fig:simu_sig_TF_pred_b400}
		\includegraphics[width=0.3\textwidth]{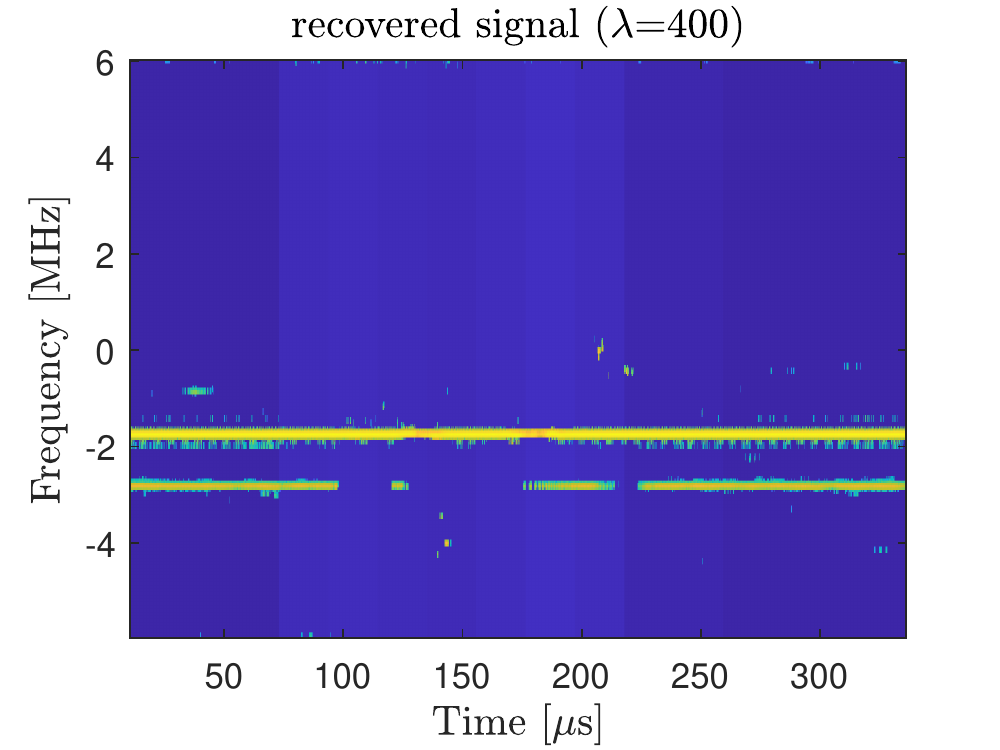}}
	\subfloat[]{\label{fig:simu_sig_fft_pred}
		\includegraphics[width=0.3\textwidth]{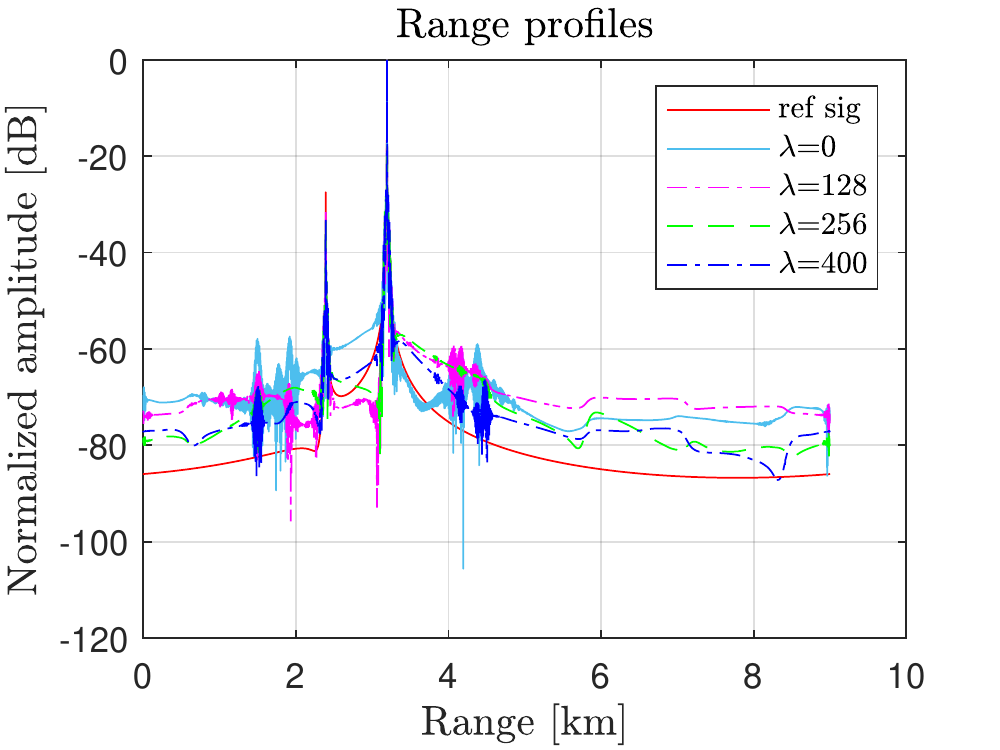}}
	\caption{Interference mitigation for simulated radar signals whose SINR is -7.91\,dB and SNR is 20\,dB. \protect\subref{fig:simu_sig_inp_RealPart} shows the acquired beat signal contaminated by mutual interferences, \protect\subref{fig:simu_sig_fft_oral_inp} its range profiles, and \protect\subref{fig:simu_sig_TF_inp} its time-frequency diagram. \protect\subref{fig:simu_sig_TF_oral} shows an interference-free reference. \protect\subref{fig:simu_sig_TF_pred_b0}-\protect\subref{fig:simu_sig_TF_pred_b400} the results of interference mitigation of optimal CV-FCN trained using one eighth of the dataset with prior-guided loss function. \protect\subref{fig:simu_sig_fft_pred} displays the corresponding range profiles obtained after interference mitigation.}
	\label{fig:tfmap}
\end{figure*}

To facilitate comparison of the regularization effect of the prior knowledge on the size of training dataset, the SINRs of recovered signals with $\lambda=400$ in Fig.~\ref{fig:reduce_dataset}\subref{fig:dataset_half}-\subref{fig:dataset_eighth} are shown together in Fig.~\ref{fig:data_reduction}. It is clear that reducing the size of the training dataset results in performance degradation over a wide range of SINRs of input signals. But with the additional prior information offered by the $L_{2,1}$ norm, the CV-FCNs trained on smaller datasets achieve comparable performance as that on the full dataset in the low SINR scenarios, even when reducing the size of the training dataset to one eighth. This is because that when the training dataset is small, the features extracted by the network using the MSE as a loss function would be insufficient for interference mitigation; thus, the network's performance would worsen. Incorporating the prior information offered by $L_{2,1}$ norm is helpful to guide and improve the features extracted for interference mitigation, compensating for the effect of data shortage. Therefore, the proposal interference mitigation approach is attractive for small data learning by introducing prior knowledge.

To demonstrate the performance of the CV-FCN (Model A) trained with 1/8 of the full data (i.e., 540 samples), Fig. \ref{fig:tfmap} illustrates the interference mitigation results of an interfered beat signal of two point targets.
Due to strong interferences between 100\,$\mu s$ and 250\,$\mu s$ in the beat signal, the weak target is almost immersed in the raised noise floor (see Fig. \ref{fig:tfmap}\subref{fig:simu_sig_inp_RealPart}-\subref{fig:simu_sig_fft_oral_inp}). After being processed with the CV-FCN obtained with $\lambda=0$, the interferences and noise are significantly suppressed, but some residual interference components are still observed (Fig.~\ref{fig:tfmap}\subref{fig:simu_sig_TF_inp} and \subref{fig:simu_sig_TF_pred_b0}).
With the increase of $\lambda$, the residual interferences and noise are further mitigated (see Fig. \ref{fig:tfmap}\subref{fig:simu_sig_TF_pred_b128}-\subref{fig:simu_sig_TF_pred_b400}), and consequently the noise floor of the range profile decreases as well (Fig.~\ref{fig:tfmap}\subref{fig:simu_sig_fft_pred}), which would improve the probability of target detection. 

Therefore, by tunning the hyper-parameter $\lambda$, the prior information characterized by $L_{2,1}$ norm can enforce the CV-FCN to extract meaningful features for interference mitigation faster during training, thus accelerating the convergence rate of training. Moreover, by incorporating the prior information, the proposed CV-FCNs can be trained with a smaller dataset, which is attractive for interference mitigation problems as it is generally very difficult to acquire labeled real radar data in practice, especially for dynamic scenarios.    

\begin{figure}[!t]
	\centering
	\vspace{-4mm}
	\includegraphics[width=0.48\textwidth]{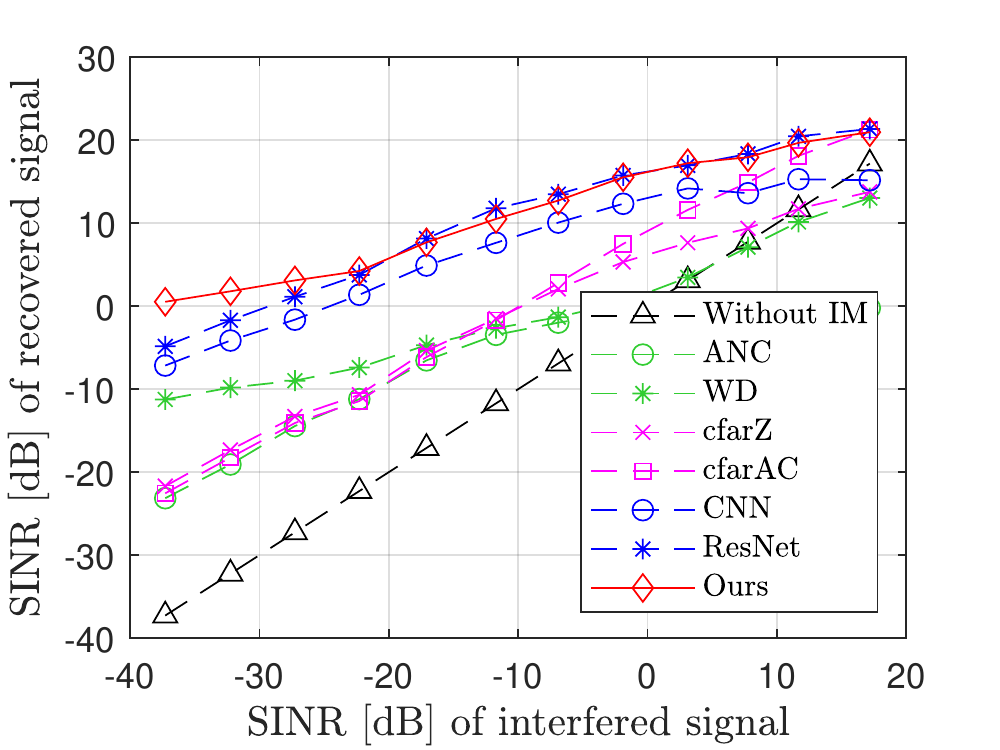}
	\caption{The performance comparison of our proposed CV-FCN based IM approach with state-of-the-art techniques} 
	\label{fig:Comparison Experiment}
\end{figure}

\subsection{Comparative Analysis with Other Techniques}

\begin{figure*}[!t]
	\centering
	\vspace{-4mm}
	\subfloat[]{\label{fig:chimney_sig_inp_RealPart}
		\includegraphics[width=0.3\textwidth]{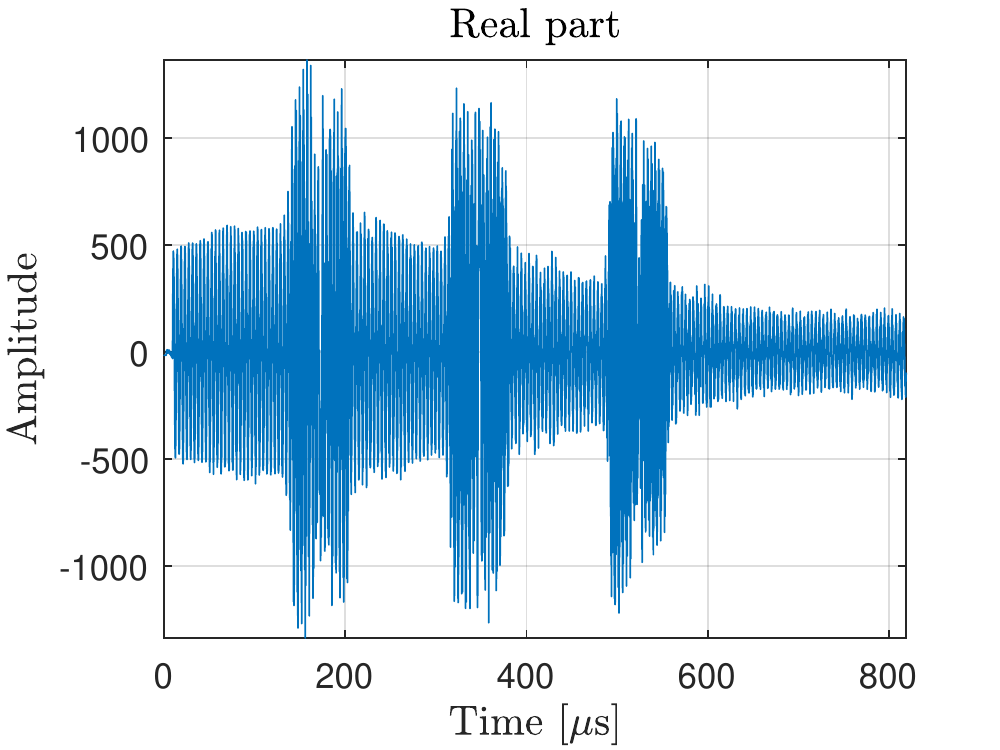}}
	\subfloat[]{\label{fig:chimney_sig_fft_inp}
		\includegraphics[width=0.3\textwidth]{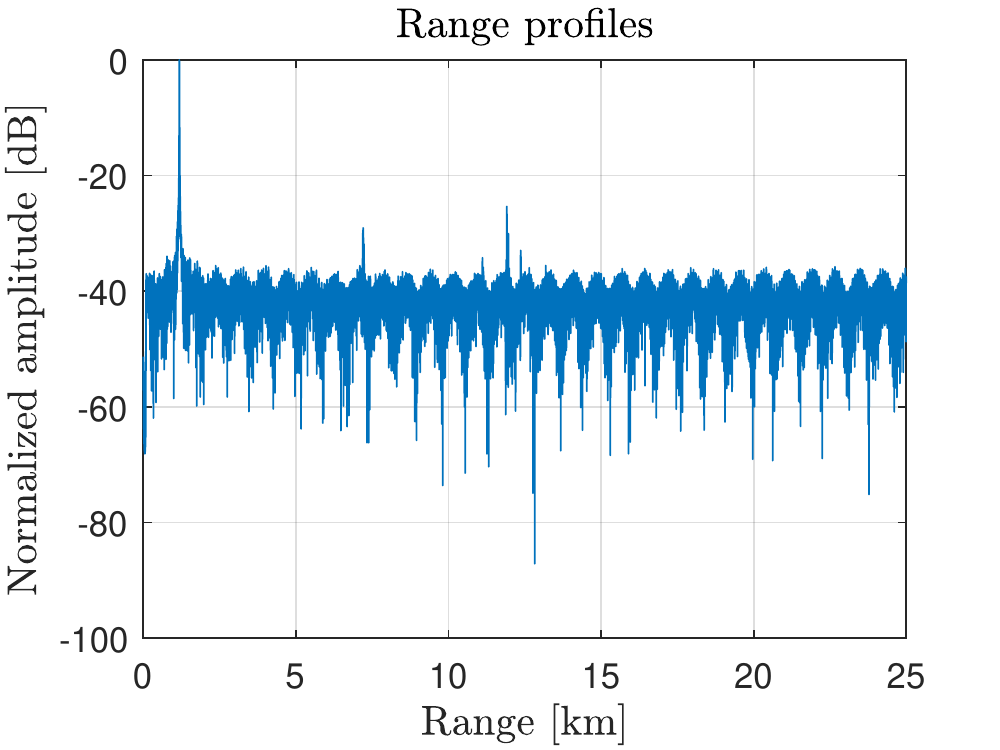}}
	\subfloat[]{\label{fig:chimney_sig_TF_inp}
		\includegraphics[width=0.3\textwidth]{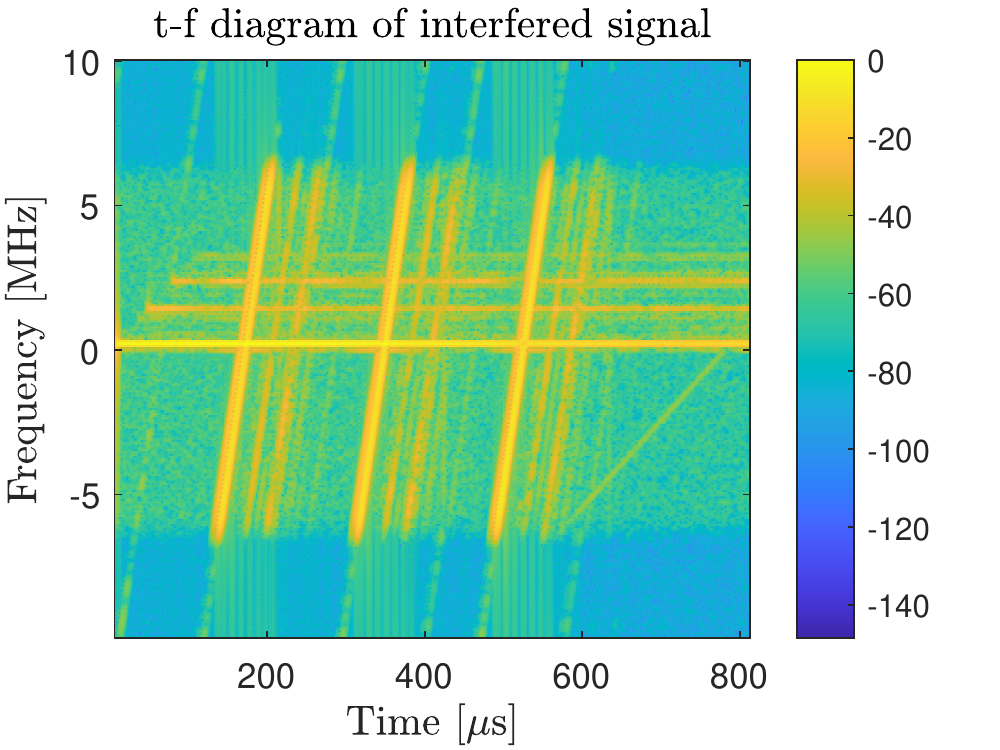}}
	
	\vspace{-3mm}
	\subfloat[]{\label{fig:chimney_sig_TF_pred_real}
		\includegraphics[width=0.3\textwidth]{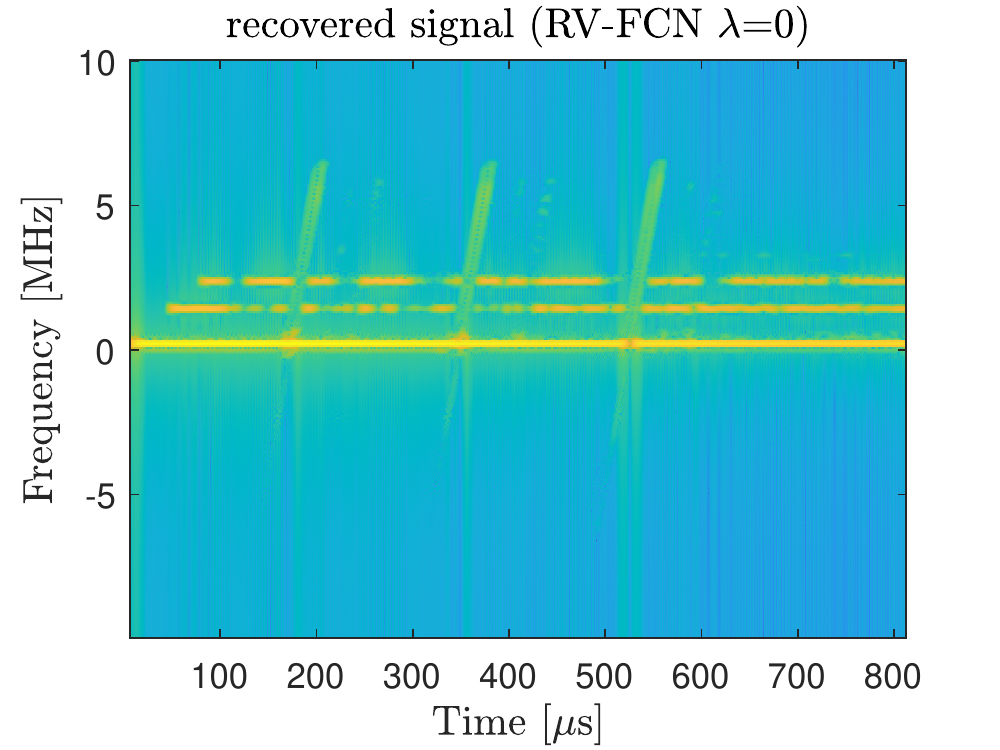}}
	\subfloat[]{\label{fig:chimney_sig_TF_pred_b_0}
		\includegraphics[width=0.3\textwidth]{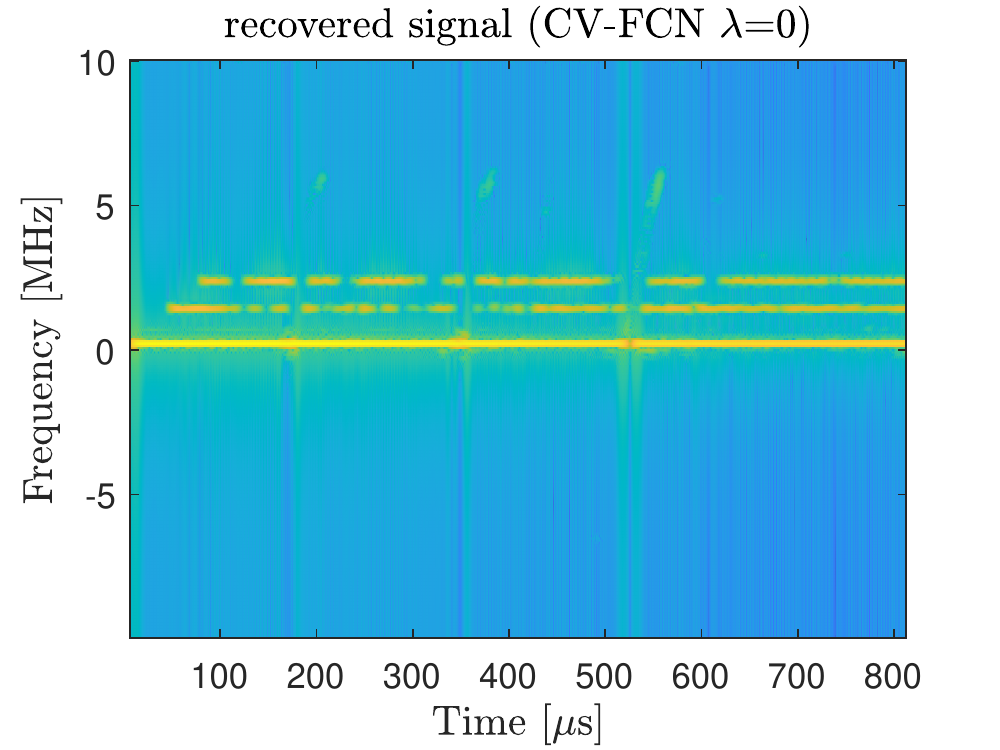}}
	\subfloat[]{\label{fig:chimney_sig_TF_pred_b_128}
		\includegraphics[width=0.3\textwidth]{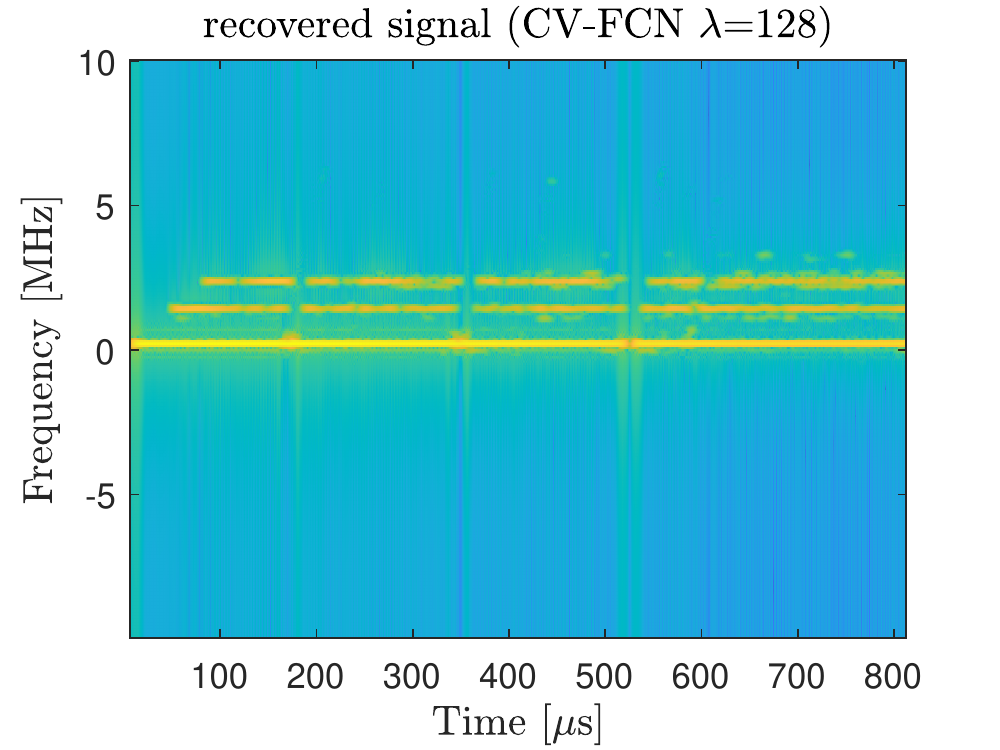}}
	
	\vspace{-3mm}
	\subfloat[]{\label{fig:chimney_sig_TF_pred_b_256}
		\includegraphics[width=0.3\textwidth]{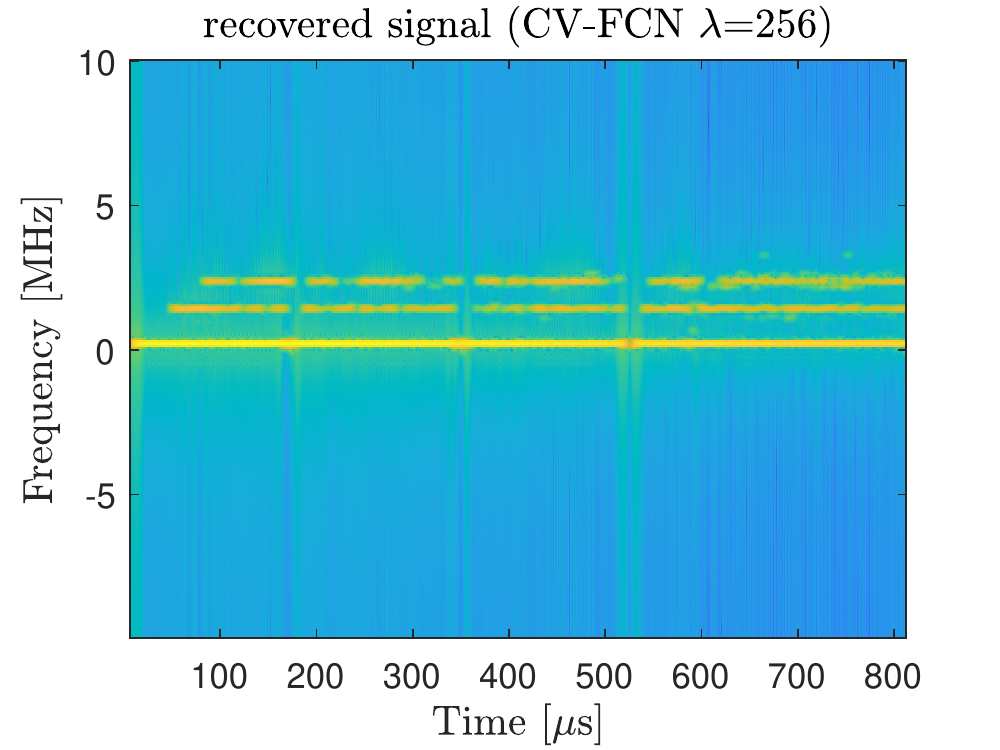}}
	\subfloat[]{\label{fig:chimney_sig_TF_pred_b_400}
		\includegraphics[width=0.3\textwidth]{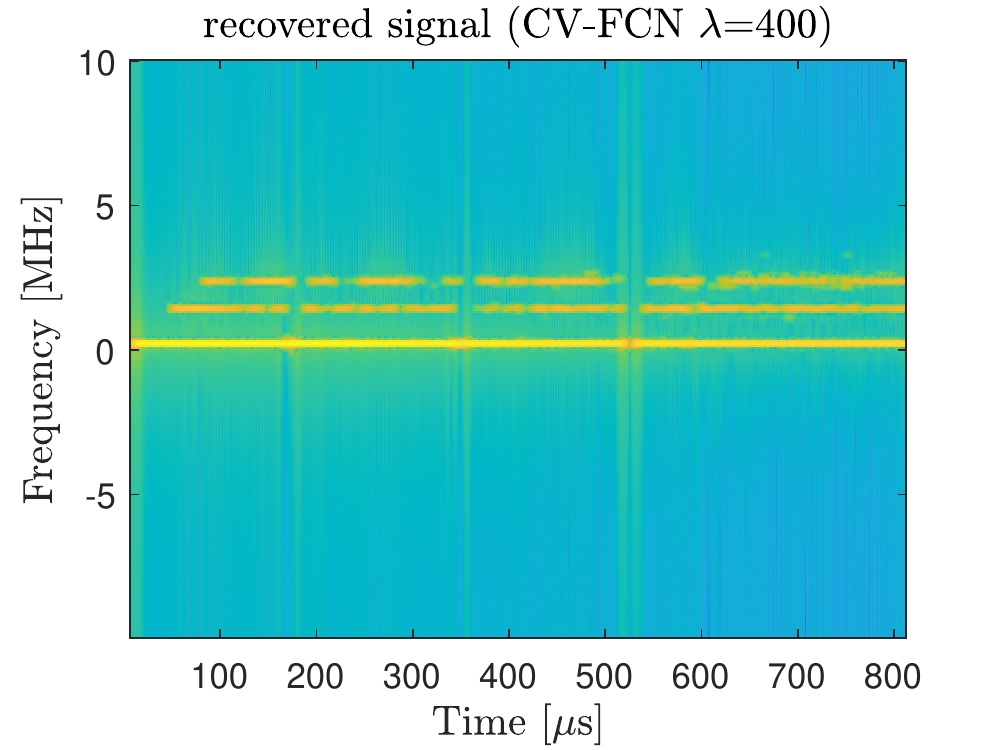}}
	\subfloat[]{\label{fig:chimney_sig_fft_pred}
		\includegraphics[width=0.3\textwidth]{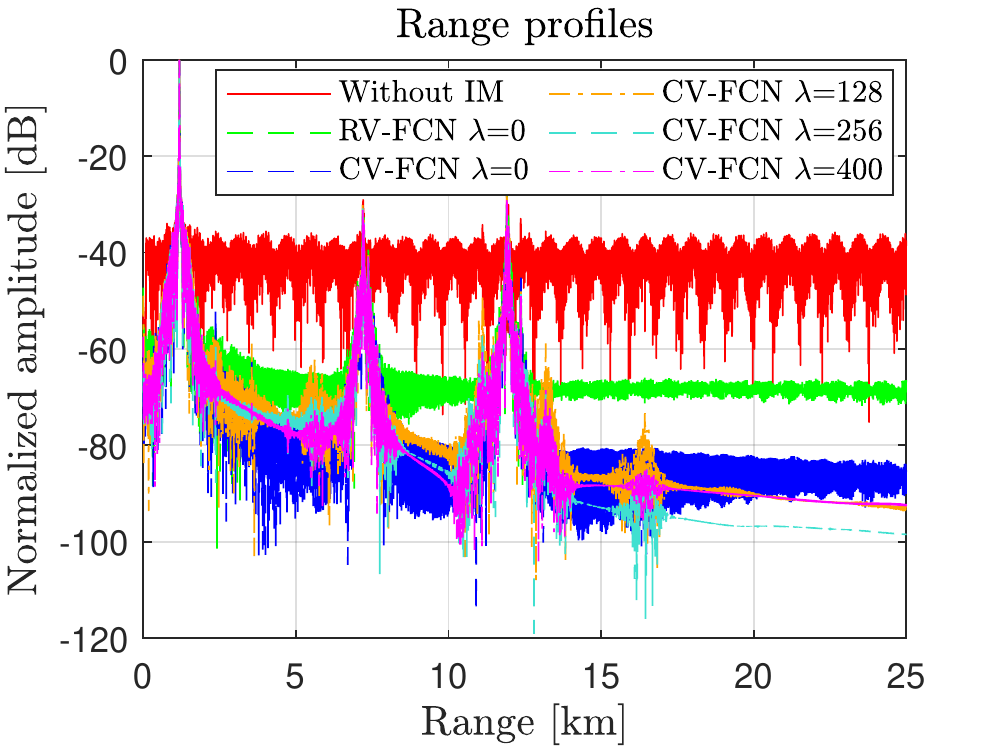}} 
	\caption{Interference mitigation for measured radar signals contaminated by the interferences from chimneys. \protect\subref{fig:chimney_sig_inp_RealPart} shows the acquired beat signal contaminated by mutual interferences \protect\subref{fig:chimney_sig_fft_inp} its range profiles and \protect\subref{fig:chimney_sig_TF_inp} its time-frequency diagram. \protect\subref{fig:chimney_sig_TF_pred_real} the results of interference mitigation of optimal RV-FCN trained with MSE. \protect\subref{fig:chimney_sig_TF_pred_b_0}-\protect\subref{fig:chimney_sig_TF_pred_b_400} the results of interference mitigation of optimal CV-FCN trained using only simulated radar signals with prior-guided loss function. \protect\subref{fig:chimney_sig_fft_pred} displays the corresponding range profiles obtained after interference mitigation. }
	\label{fig_chimney}
\end{figure*}

\begin{figure*}[!t]
	\centering
	\vspace{-4mm}
	\subfloat[]{\label{fig:sig_TF_inp_A13}
		\includegraphics[width=0.3\textwidth]{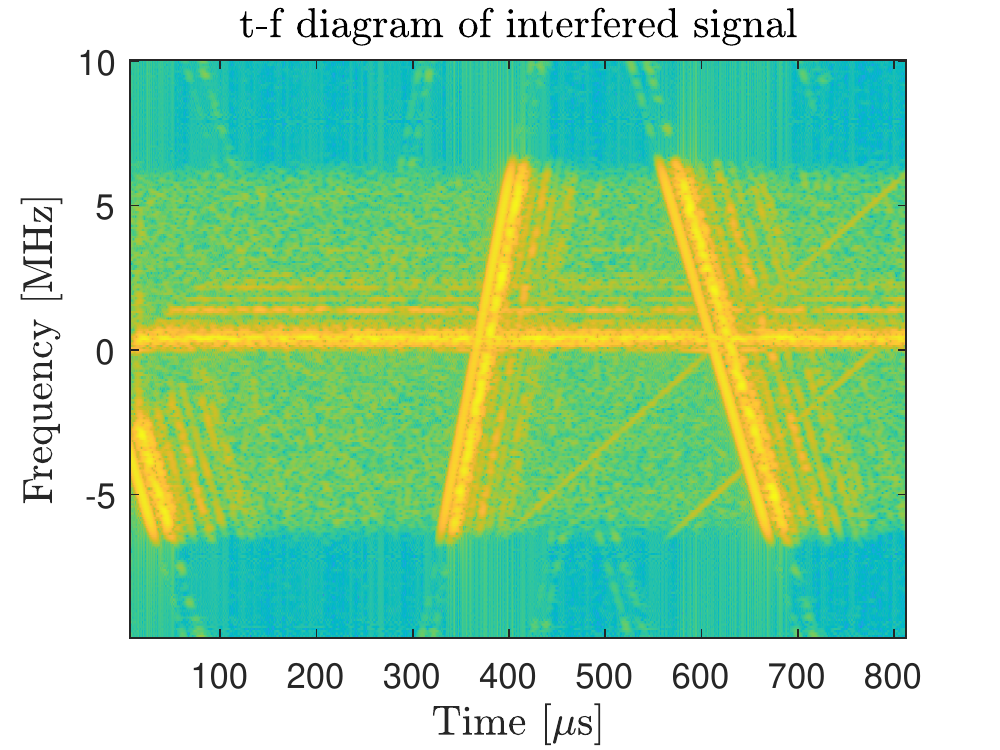}}
	\subfloat[]{\label{fig:sig_TF_inp_straat}
		\includegraphics[width=0.3\textwidth]{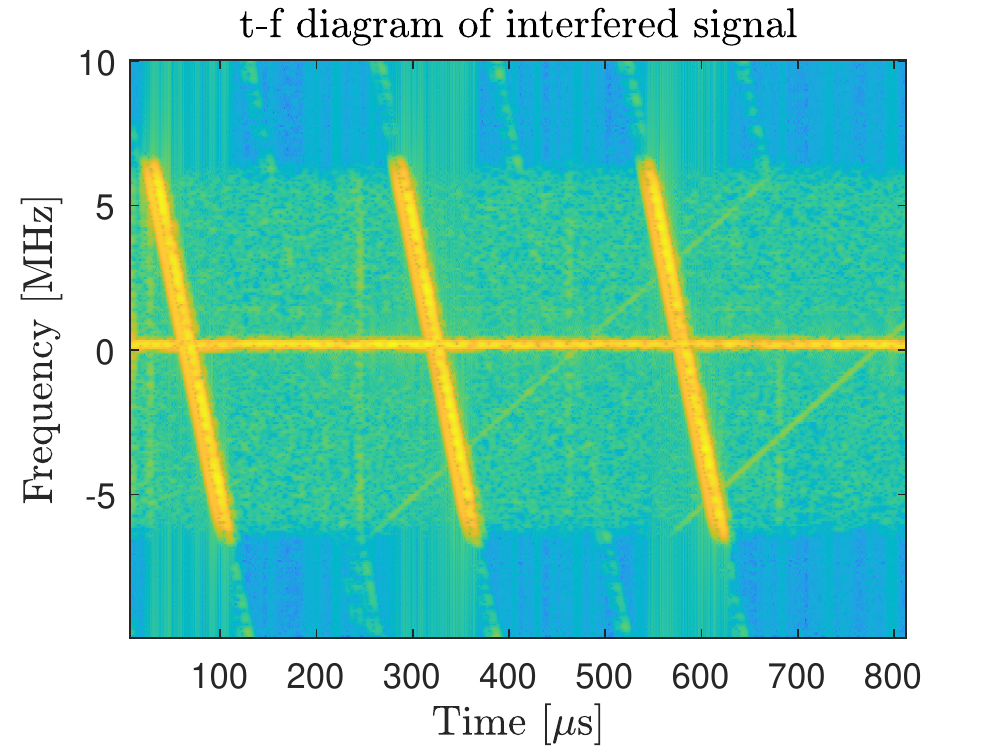}}
	\subfloat[]{\label{fig:sig_TF_inp_windmill}
		\includegraphics[width=0.3\textwidth]{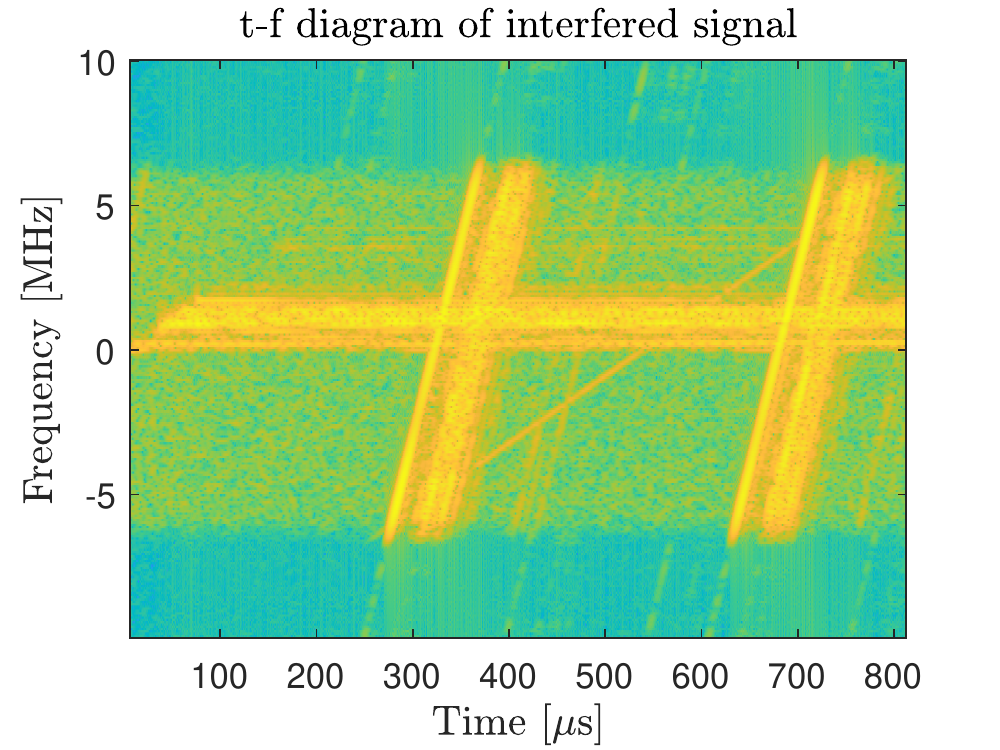}}
	
	\vspace{-3mm}
	\subfloat[]{\label{fig:sig_TF_pred_A13_real}
		\includegraphics[width=0.3\textwidth]{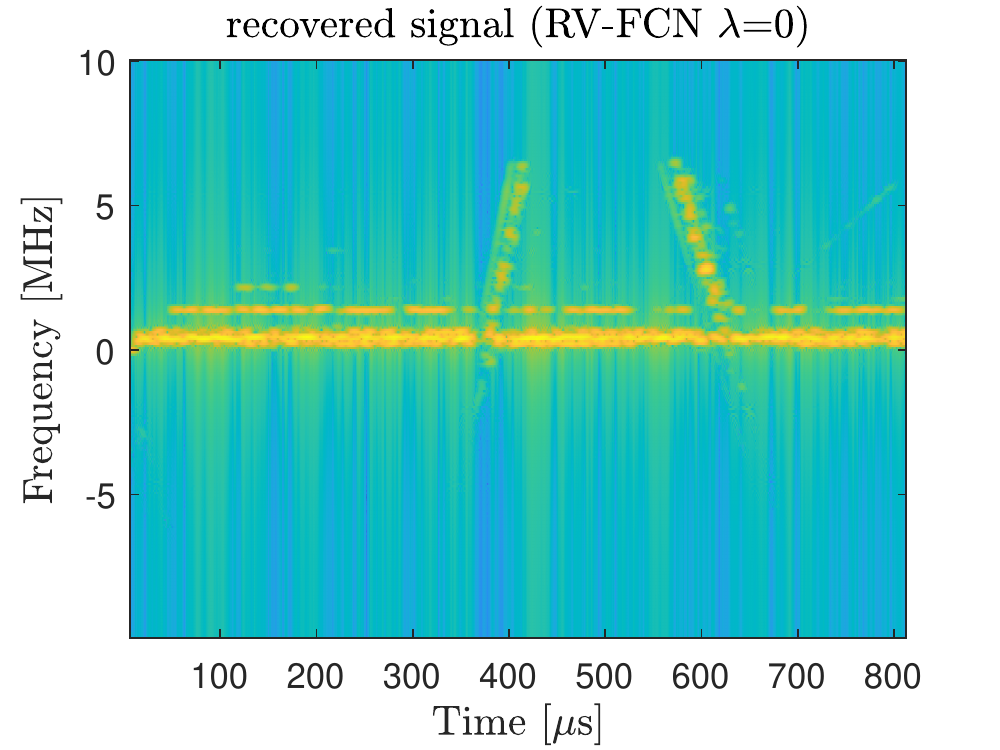}}
	\subfloat[]{\label{fig:sig_TF_pred_straat_real}
		\includegraphics[width=0.3\textwidth]{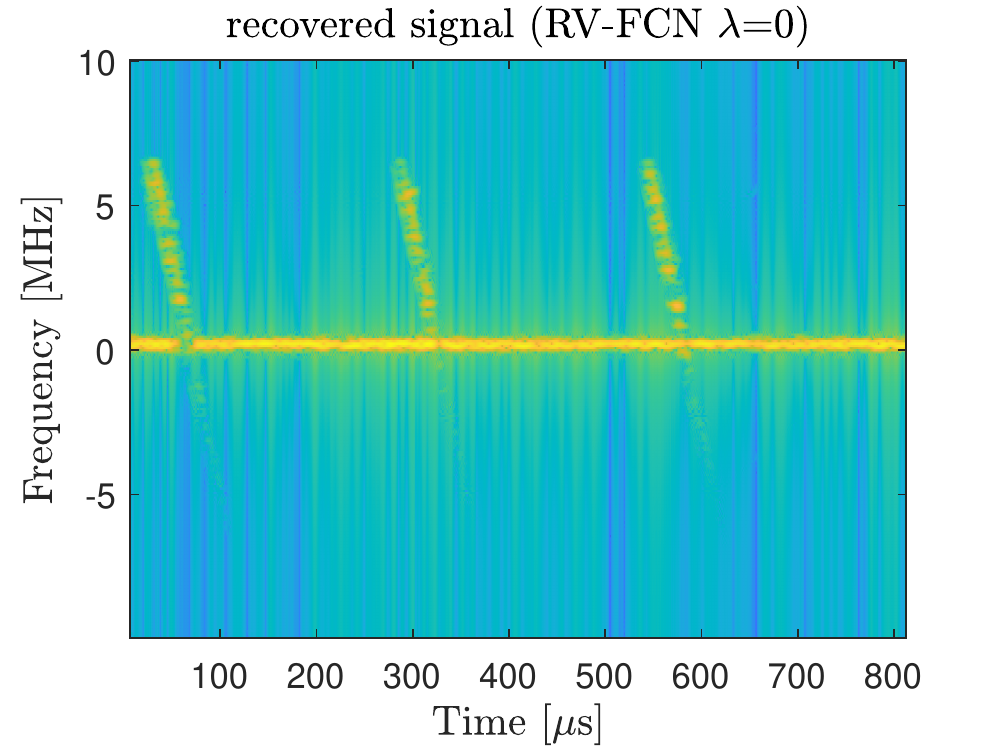}}
	\subfloat[]{\label{fig:sig_TF_pred_windmill_real}
		\includegraphics[width=0.3\textwidth]{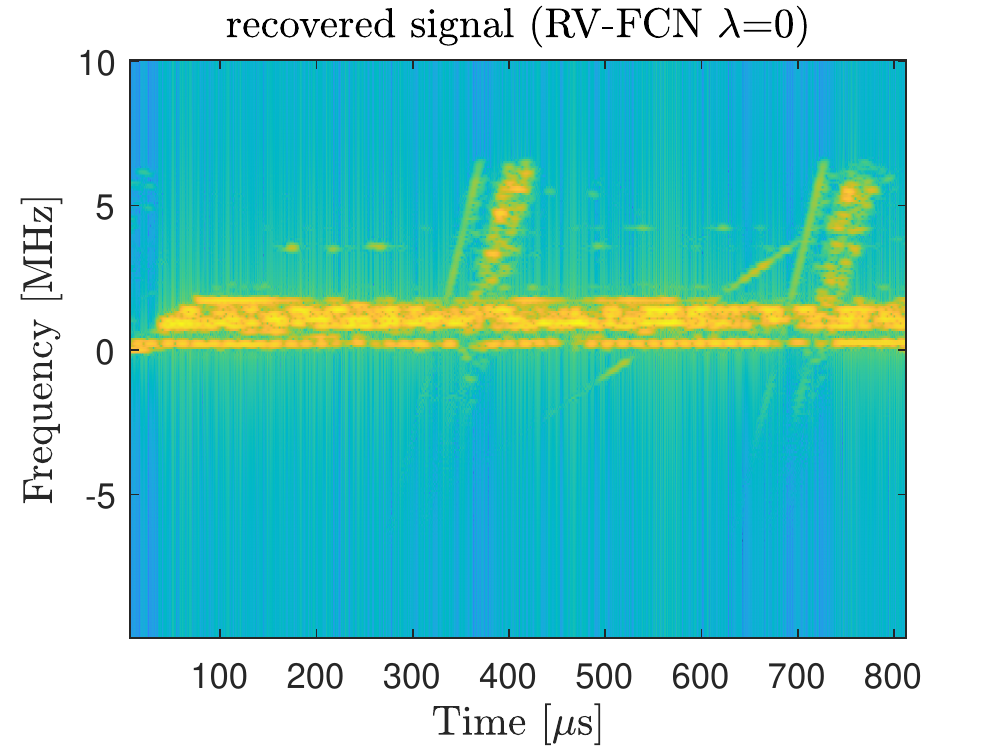}}
	
	\vspace{-3mm}
	\subfloat[]{\label{fig:sig_TF_pred_A13_b_0}
		\includegraphics[width=0.3\textwidth]{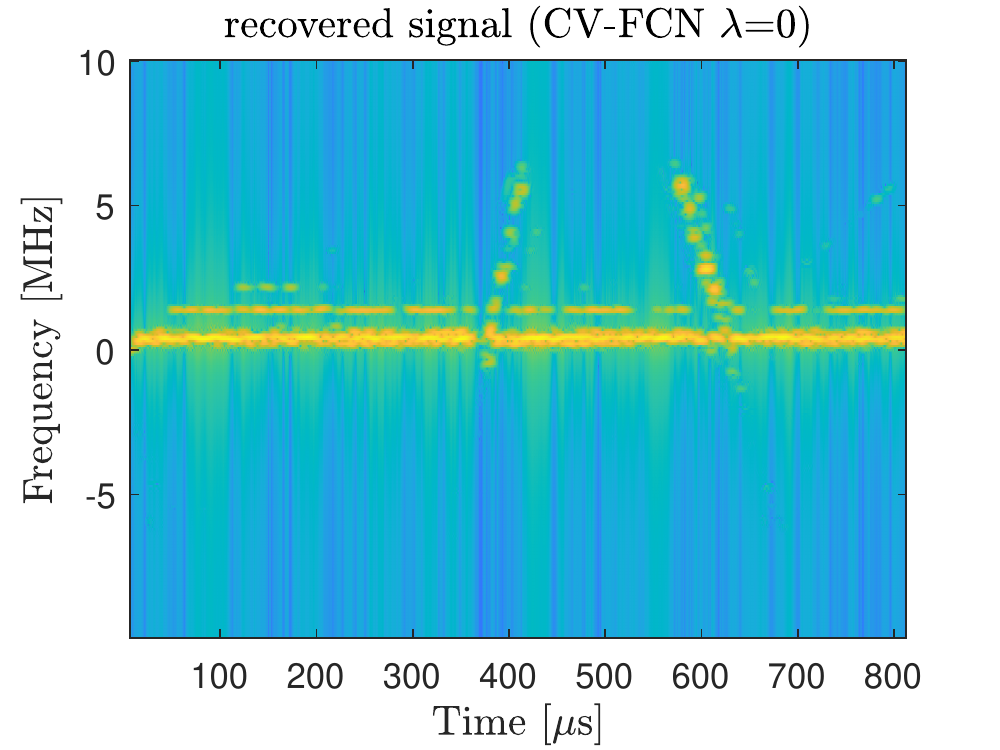}}
	\subfloat[]{\label{fig:sig_TF_pred_straat_b_0}
		\includegraphics[width=0.3\textwidth]{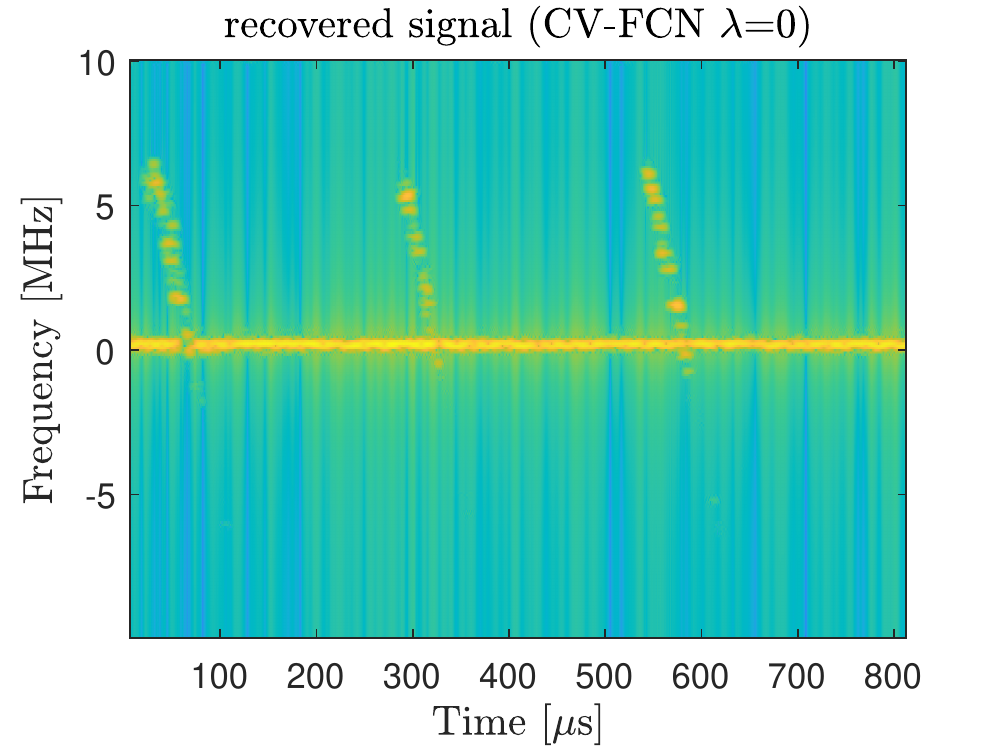}}
	\subfloat[]{\label{fig:sig_TF_pred_windmill_b_0}
		\includegraphics[width=0.3\textwidth]{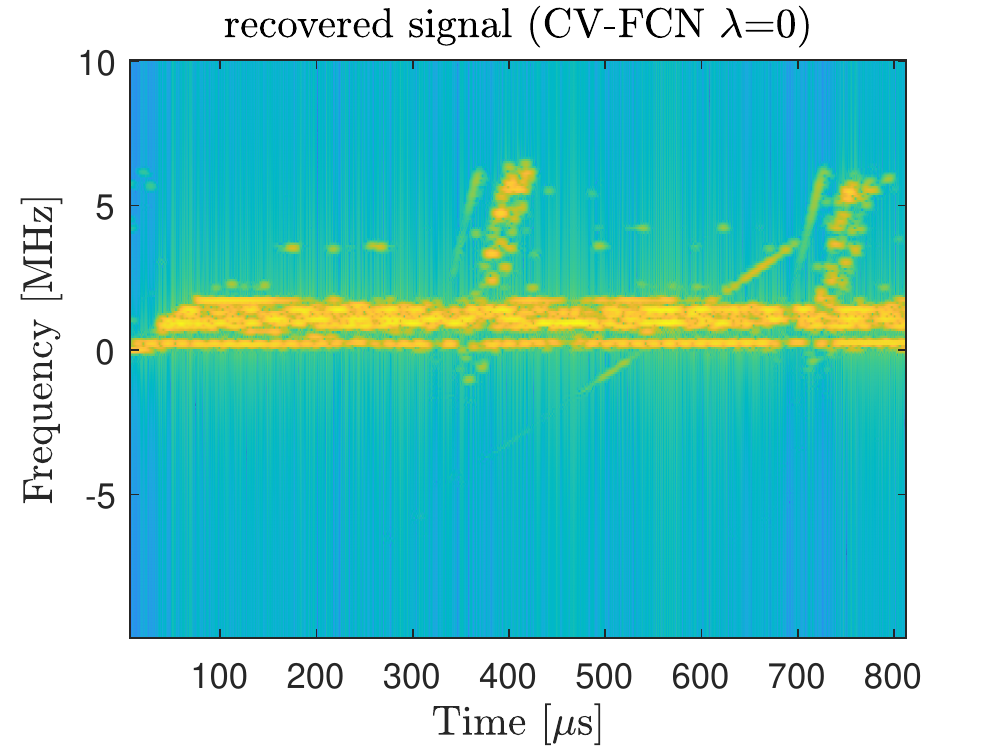}}
	
	\vspace{-3mm}
	\subfloat[]{\label{fig:sig_TF_pred_A13_b_400}
		\includegraphics[width=0.3\textwidth]{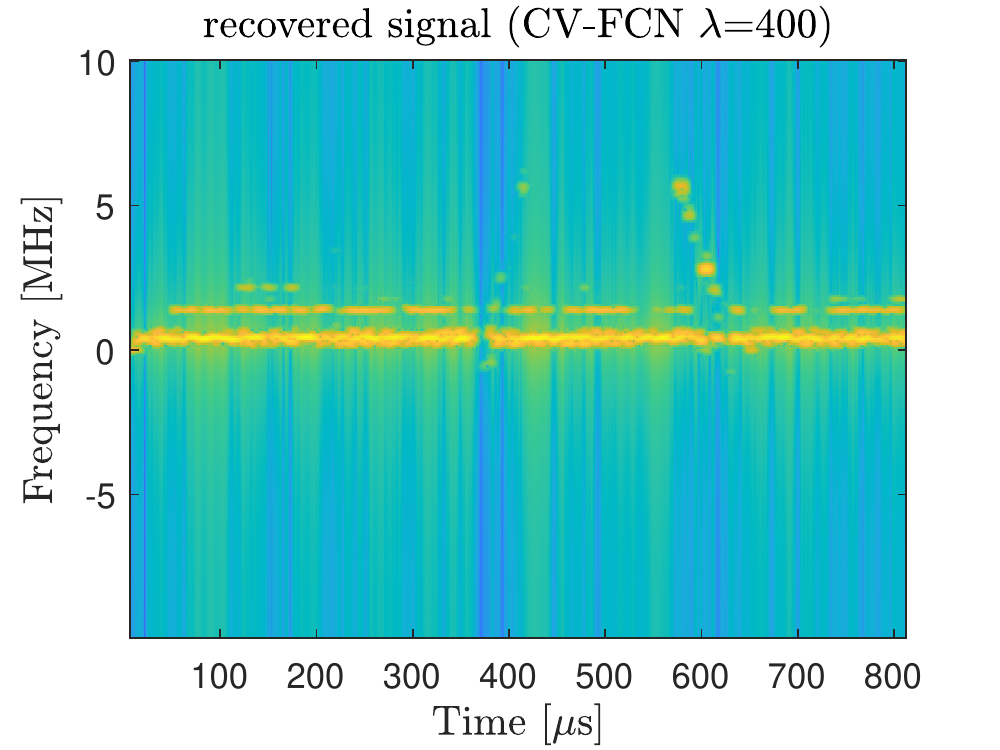}}
	\subfloat[]{\label{fig:sig_TF_pred_straat_b_400}
		\includegraphics[width=0.3\textwidth]{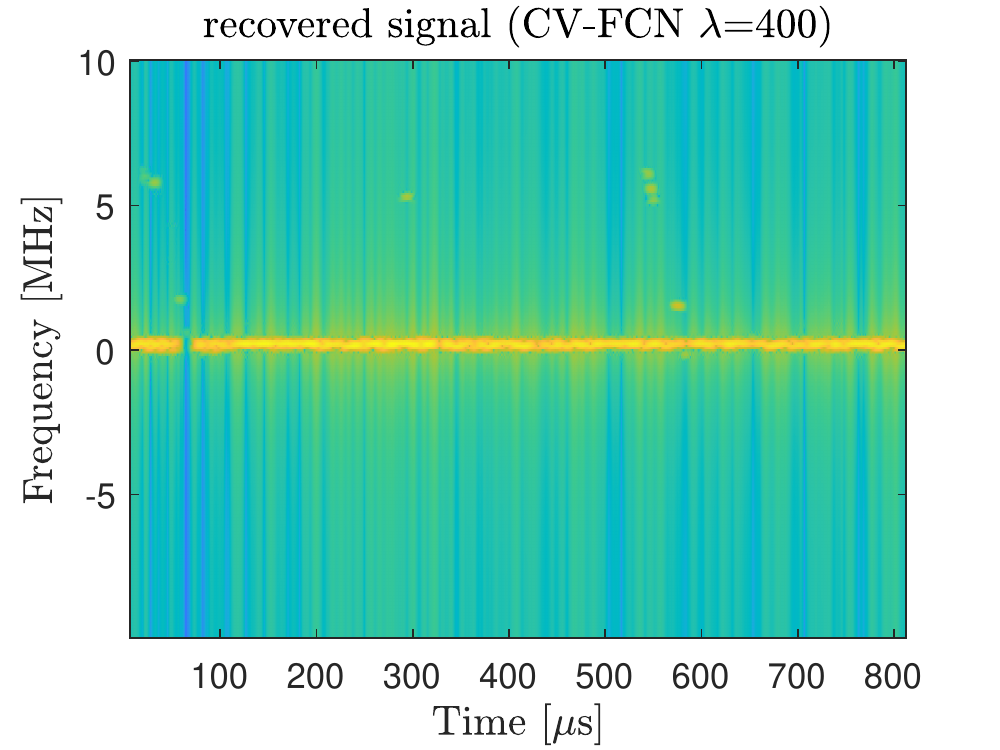}}
	\subfloat[]{\label{fig:sig_TF_pred_windmill_b_400}
		\includegraphics[width=0.3\textwidth]{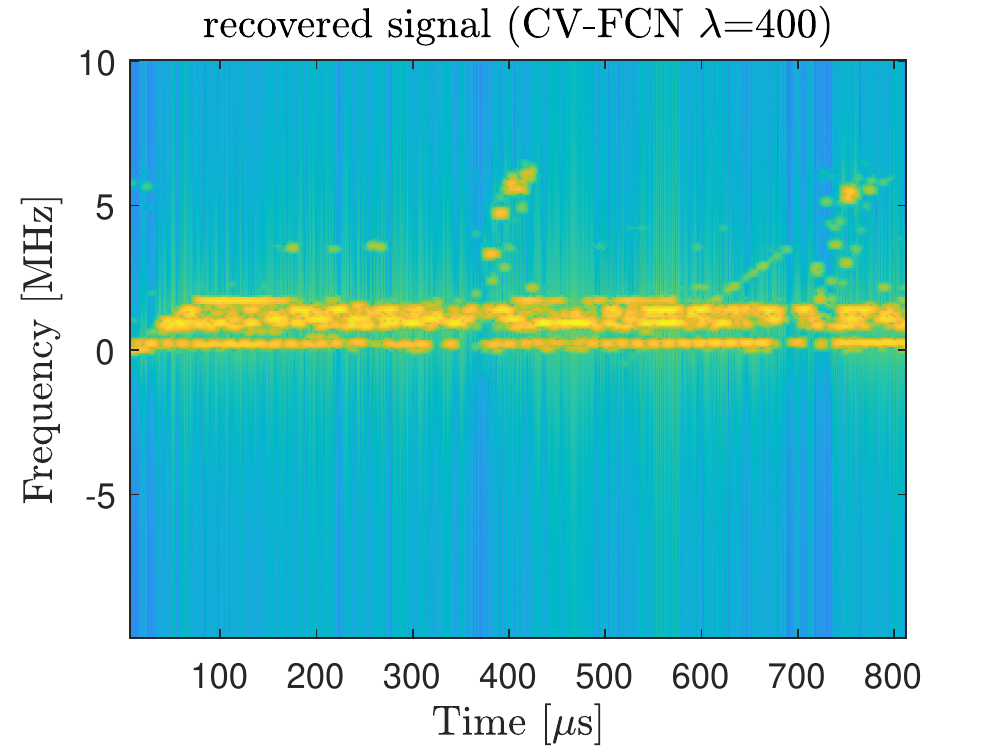}}
	
	\caption{Interference mitigation for measured radar signals in three scenarios (A13, windmill and street from left to right). \protect\subref{fig:sig_TF_inp_A13}-\protect\subref{fig:sig_TF_inp_straat} shows the time-frequency diagram of acquired beat signal contaminated by mutual interferences. \protect\subref{fig:sig_TF_pred_A13_b_0}-\protect\subref{fig:sig_TF_pred_straat_b_0} the results of interference mitigation of optimal CV-FCN trained with prior-guided loss function where b is 0. \protect\subref{fig:sig_TF_pred_A13_b_400}-\protect\subref{fig:sig_TF_pred_straat_b_400} the results of interference mitigation of optimal CV-FCN trained with prior-guided loss function where b is 400.}
	\label{fig:measured_results}
\end{figure*}
The performance of our proposed approach is compared with several state-of-the-art interference mitigation methods, including traditional signal processing approaches such as the Wavelet Denoising (WD) based method\cite{Wavelet_Denoising}, Adaptive Noise Canceller (ANC) \cite{ANC}, CFAR-Z and CFAR-AC \cite{CFAR}, and DL-based approaches such as CNN-based method \cite{CNN} and ResNet-based method \cite{resnet}.
We used the simulated radar signals for test and quantitatively evaluated the performance of different methods.
Fig. \ref{fig:Comparison Experiment} shows the SINRs of the obtained signals after interference mitigation with different approaches. 
Due to the ResNet was designed to process the SAR images, in our experiments, the number of filters of the ResNet is set to half of that in \cite{resnet} for FMCW radar signal processing.
The CV-FCN is trained using the prior-guided loss function where $\lambda=400$ and the CNN and ResNet are trained using the MSE loss function.

The comparative results show that our proposed CV-FCN based prior-guided IM approach is obviously better than other methods. 
Specifically, the cfarZ and cfarAC use constant false alarm rate (CFAR) to detect the interference components of acquired beat signals in the time-frequency spectrum. The detection accuracy is determined by the selected parameters.
Then cfarZ uses zeroing to mitigate detected interferences, which naturally removes targets' beat signals at the same time. 
Different from cfarZ, cfarAC uses amplitude correction (AC) to reconstruct the beat signals removed by zeroing, which shows better performance than cfarZ. 
Besides, WD method can extract and remove the interferences in the wavelet domain, which shows a good performance in low SINR scenarios. 
In the ANC method, the negative half of the FFT spectrum is used as the input of its reference channel, and the filtering step size is manually adjusted. 
As described, the performance of the above traditional signal processing methods depends on a proper selection of a few manually adjustable parameters. 
Over a wide range of the SINR variations, their performance is not good as the selected DL-based methods\footnote{In principle, the range of  SINRs of input signals can be divided into a few small segments. By tuning the related parameters of conventional methods over each small range of SINR of input signals, they could outperform the DL-based methods. However, selecting such a set of parameters is nontrivial in practice.}.

On the other hand, our proposed CV-FCN achieves better performance expecially in low SINR scenarios with only 12\% the number of total parameters of the ResNet. 
The effect of residual connection has been verified and discussed in Section V-B, which is not suitable for extracting the feature of targets' beat signals. 
The superior performance compared with other NNs shows the advantage of network architecture optimization by grid search, complex-valued representation in radar signal processing, and the prior feature offered by $L_{2,1}$ norm.

\section{Measuremnt Results}
In this section, the radar signals measured as described in Section IV-B are used to verify the generalization of our proposed prior-guided CV-FCN based IM approach.

We consider the scene of industrial chimneys.
Due to the limitation of the experimental condition, the clean reference signal cannot be obtained.
The qualitative results, including the signal waveforms in the time-domain, the $t$-$f$ diagrams, and range profiles of beat signals, are shown in Fig. \ref{fig_chimney}.
As shown in Fig. \ref{fig_chimney}\subref{fig:chimney_sig_inp_RealPart}, three large pulses can be observed in the received radar signal, which is caused by the strong interferences. 
Then the interference-contaminated beat signal leads to a range profile with significantly increased noise floor, and the two weaker targets cannot be detected (see Fig. \ref{fig_chimney}\subref{fig:chimney_sig_fft_inp}). 
The $t$-$f$ spectrum of the beat signal is computed through the STFT algorithm, where the parameters setting is the same as simulated signals. 
One can see from Fig. \ref{fig_chimney}\subref{fig:chimney_sig_TF_inp} that the interferences exhibit as three inclined thick lines in the $t$-$f$ spectrum. 

To overcome the missed detection of targets caused by the strong interferences, the optimal RV-FCN and CV-FCN obtained in Section V-B are used to suppress the interference components in measured radar signals.
The network was trained using the dataset including only simulated radar signals, and the prior-guided loss function is used for training.

The $t$-$f$ map of the recovered signal processed by the RV-FCN is shown in Fig. \ref{fig_chimney}\subref{fig:chimney_sig_TF_pred_real}, the interferences are entirely removed in the negative frequency, but there are still residual interference components mixed with the desired spectrum of targets in the positive frequency. 
This can be explained by the fact that the beat signals mixed with interferences and noise are relatively more difficult for the network to extract their features. 
In contrast, a more complete interference mitigation performance can be seen in Fig. \ref{fig_chimney}\subref{fig:chimney_sig_TF_pred_b_0}, which shows the better generalization performance of complex-valued networks.
Furthermore, with the value of $\lambda$ increases, the residual interference components and noise are removed as shown in Fig. \ref{fig_chimney}\subref{fig:chimney_sig_TF_pred_b_0} to \subref{fig:chimney_sig_TF_pred_b_400}.
After interference mitigation, three peaks representing the objects can be clearly seen in the range profiles as shown in Fig. \ref{fig_chimney}\subref{fig:chimney_sig_fft_pred}, and the CV-FCN offers a lower noise floor than the RV-FCN. 
With the proportion of $L_{2,1}$ norm gradually increases, the noise floor is further decreased, which would help to improve the target detection probability.

We also applied the RV-FCN and our proposed approach to the measured radar signals collected in the other two scenes for interference mitigation (a rotating wind turbine and a street with moving cars). 
The $t$-$f$ maps of interference contaminated beat signals and recovered radar signals are shown in Fig. \ref{fig:measured_results}.
Similarly, a better IM performance is obtained by the CV-FCN, but there are still interference components in the positive half of the STFT spectrum.
By increasing the value of $\lambda$, this problem can be solved. 
The residual interference components and noise are obliterated, and the desired spectra related to targets are recovered.

The experimental results on measured radar signals collected in various real-world scenes have shown a better generalization performance of complex-valued networks.
Besides, we can see the effectiveness of the prior-guided loss function in helping the neural networks to remove the residual interferences and noise in measured radar signals. 
Therefore, the proposed CV-FCN based prior-guided interference mitigation approach can be better applied in reality.

\section{Conclusion}

In this paper, a prior-guided deep learning based interference mitigation approach has been presented for FMCW radars. 
The STFT is used to transform received radar signals to the $t$-$f$ domain so that the NNs can better extract the features of targets' beat signals and interferences. 
Then the CV-FCN is designed to deal with the complex-valued radar signals. 
Meanwhile, the prior feature is introduced as a regularization term in the training stage.

An FMCW radar interference dataset with a wide and realistic range of signal parameter variations is presented. 
The experimental results show a better interference mitigation performance with fewer parameters in low SINR scenarios offered by complex-value networks. 
Additionally, the networks can converge faster, and the size of the dataset needed for training can be reduced with the prior-guided loss function. 
Compared to the well-known traditional and DL-based interference mitigation techniques, the proposed approach achieves the state-of-the-art in SINR based performance comparison. Finally, the qualitative results on the measured radar signals show its excellent generalization. In future work, we aim to design an optimization algorithm to adjust the value of hyper-parameter $\lambda$ automatically in the prior-guided loss function. 

\section*{Acknowledgment}

The authors would like to thank F. van der Zwan from TU Delft for his help for experimental measurements and providing the data. 

\ifCLASSOPTIONcaptionsoff
  \newpage
\fi



\bibliographystyle{hieeetr}
\bibliography{IEEEabrv,ref.bib}
\end{document}